%% file: ms.tex
\documentclass[a4paper,fleqn,usenatbib]{mnras}

\usepackage{deluxetable}
\usepackage{newtxtext,newtxmath}
\usepackage[T1]{fontenc}
\usepackage{ae,aecompl}
\usepackage{amssymb, amsmath}
\usepackage{graphicx}
\usepackage{natbib}
\usepackage{url}
\usepackage{hyperref}
\usepackage{float}
\usepackage[usenames, dvipsnames]{color}

\hypersetup{colorlinks=true,
            citecolor=MidnightBlue,
            linkcolor=MidnightBlue,
            filecolor=magenta,      
            urlcolor=cyan}
\DeclareGraphicsExtensions{.pdf,.png,.jpg}

\def\arcsec{{\prime\prime}}

\def\aj{{AJ}}

\def\apj{{ApJ}}
\def\apjs{{ApJS}}
\def\asec{$^{\prime\prime}$}

\def\lax{{$\mathrel{\hbox{\rlap{\hbox{\lower4pt\hbox{${\sim}$}}}\hbox{$<$}}}$}}
\def\gax{{$\mathrel{\hbox{\rlap{\hbox{\lower4pt\hbox{${\sim}$}}}\hbox{$>$}}}$}}
\def\simlt{\lower.5ex\hbox{$\; \buildrel < \over {\sim} \;$}}
\def\simgt{\lower.5ex\hbox{$\; \buildrel > \over {\sim} \;$}}

\def\mnras{{MNRAS}}
\def\nat{{Nature}}
\def\pasp{{PASP}}

\def\sb{mag~arcsec$^{-2}$}
 
\def\msun{$M_\odot$}

\def\etal{{\ et al.~}}

\def\ser{{S\'{e}rsic\ }}
\def\redm{\texttt{redMaPPer}}
\def\cmodel{\texttt{cModel}}
\def\rbcg{\texttt{cenHighMh}}
\def\nbcg{\texttt{cenLowMh}}

\def\mstar{{$M_{\star}$}}
\def\mhalo{{$M_{\mathrm{200b}}$}}
\def\logms{{$\log_{10} (M_{\star}/M_{\odot})$}}
\def\logmh{{$\log_{10} (M_{\mathrm{200b}}/M_{\odot})$}}
\def\logmhalo{{$\log_{10} (M_{\mathrm{200b}}/M_{\odot})$}}

\def\minn{{$M_{\star,10\mathrm{kpc}}$}}
\def\meff{{$M_{\star,15\mathrm{kpc}}$}} 
\def\mtot{{$M_{\star,100\mathrm{kpc}}$}}

\def\mmax{{$M_{\star,\mathrm{Max}}$}}
\def\mgama{{$M_{\star,\mathrm{GAMA}}$}}

\def\logminn{{$\log_{10} (M_{\star,10\mathrm{kpc}}/M_{\odot})$}}
\def\logmtot{{$\log_{10} (M_{\star,100\mathrm{kpc}}/M_{\odot})$}}

\def\logmgama{{$\log_{10} (M_{\star,\mathrm{GAMA}}/M_{\odot})$}}

\def\m2l{{$M_{\star}/L_{\star}$}}
\def\s2n{{$\mathrm{S}/\mathrm{N}$}}
\def\mden{{$\mu_{\star}$}}

\newcommand{\xxx}[1]{\textcolor{red}{\textbf{XXX}}}

\newcommand{\song}[1]{\textcolor{cyan}{#1}}

\title[Structure and Environment of Massive Galaxies]{
       A Detection of the Environmental Dependence of the Sizes and Stellar Haloes
       of Massive Central Galaxies}
       
\author[S. Huang et al.]{
        Song Huang$^{1,2}$\thanks{E-mail: song.huang@ipmu.jp (SH)},
        Alexie Leauthaud$^{1}$,
        Jenny Greene$^{4}$,
        Kevin Bundy$^{3}$,
        \newauthor
        Yen-Ting Lin$^{5}$,
        Masayuki Tanaka$^{6}$,
        Rachel Mandelbaum$^{7}$,
        Satoshi Miyazaki$^{5,8}$,
        \newauthor
        Yutaka Komiyama$^{5,8}$
        \\
        $^{1}$Department of Astronomy and Astrophysics, University of California 
              Santa Cruz, 1156 High St., Santa Cruz, CA 95064, USA\\
        $^{2}$Kavli-IPMU, The University of Tokyo Institutes for Advanced Study, 
              the University of Tokyo (Kavli IPMU, WPI), Kashiwa 277--8583, Japan\\              
        $^{3}$UCO/Lick Observatory, University of California, Santa Cruz,
              1156 High Street, Santa Cruz, CA 95064, USA\\
        $^{4}$Department of Astrophysical Sciences, Peyton Hall,
              Princeton University, Princeton, NJ 08540, USA \\
        $^{5}$Academia Sinica Institute of Astronomy and Astrophysics, 
              P.O. Box 23--141, Taipei 10617, Taiwan\\
        $^{6}$National Astronomical Observatory of Japan, 2--21--1 Osawa, Mitaka, 
              Tokyo 181--8588, Japan\\
        $^{7}$McWilliams Center for Cosmology, Department of Physics, 
              Carnegie Mellon University, Pittsburgh, PA 15213, USA\\
        $^{8}$SOKENDAI (The Graduate University for Advanced Studies), Mitaka,
              Tokyo, 181--8588, Japan
        }   
\date{Accepted XXX. Received YYY; in original form ZZZ}        
\pubyear{2017}                                  
  
\begin{document}

\label{firstpage}
\pagerange{\pageref{firstpage}--\pageref{lastpage}}

\maketitle


\begin{abstract}
  
    We use ${\sim}100$ deg$^2$ of deep ($>28.5$ \sb{} in $i$-band), high-quality 
    (median 0.6\asec seeing) imaging data from the Hyper Suprime--Cam (HSC) survey to 
    reveal the halo mass dependence of the surface mass density profiles and outer 
    stellar envelopes of massive galaxies. 
    The $i$-band images from the HSC survey reach ${\sim}4$ magnitudes deeper than 
    Sloan Digital Sky Survey and enable us to directly trace stellar mass distributions 
    to 100 kpc without requiring stacking.  
    We conclusively show that, at fixed stellar mass, the stellar profiles of massive 
    galaxies depend on the masses of their dark matter haloes. 
    On average, massive central galaxies with \logmtot{}$>11.6$ in more massive haloes 
    at $0.3 < z < 0.5$ have shallower inner stellar mass density profiles 
    (within ${\sim}10$--$20$ kpc) and more prominent outer envelopes. 
    These differences translate into a halo mass dependence of the mass--size relation. 
    Central galaxies in haloes with \logmh{}$>14.0$ are $\sim 20$\% larger in 
    $R_{\mathrm{50}}$ at fixed \mtot{}. 
    Such dependence is also reflected in the relationship between the stellar mass 
    within 10 and 100 kpc.  
    Comparing to the mass--size relation, the \mtot{}--\minn{} relation avoids the 
    ambiguity in the definition of size, and can be straightforwardly compared with 
    simulations. 
    Our results demonstrate that, with deep images from HSC, we can quantify the 
    connection between halo mass and the outer stellar halo, which may provide new 
    constraints on the formation and assembly of massive central galaxies.
    
\end{abstract}




\section{Introduction}
    \label{sec:intro}
    
    A key discovery in the last decade has been the dramatic structural transformation 
    of massive quiescent galaxies \citep[e.g.,][]{Trujillo2006, vanDokkum2008, 
    Cimatti2008, Damjanov2009, vanderWel2011, Szomoru2012, Patel2013} from 
    $z \approx 2$ to the present day. 
    These observations suggest that the progenitors of $z{\sim} 0$ massive early-type 
    galaxies (ETGs) need to increase their effective radii ($R_{\rm e}$) by a factor 
    of 2--4 over a time span of 10 Gyrs (e.g., \citealt{Newman2012, vdWel2014}). 
    This observational result spurred the development of the `two-phase' formation
    scenario for massive ETGs (e.g., \citealt{Oser2010, Oser2012}),
    in which galaxies form a compact central region at $z\sim 2$ through highly 
    dissipative processes (e.g., gas-rich mergers or cold gas-accretion;
    \citealt{Hopkins2008, Dekel2009}). 
    They subsequently assemble extended stellar haloes via dry mergers 
    \citep[e.g.,][]{Naab2006, Khochfar2006, Oser2010, Oser2012}, which can cause 
    significant size growth at late times. 
    An alternative explanation for size growth, progenitor bias, hypothesizes that 
    larger ETGs were quenched more recently; but this explanation is still under 
    active debate (e.g., \citealt{Newman2012, Carollo2013, Poggianti2013, Belli2015,
    Keating2015, Fagioli2016}). 
    
    There have been multiple observational attempts to test the two-phase 
    formation scenario using galaxies at low redshift, by investigating surface 
    brightness or mass density profiles (e.g., \citealt{Huang2013a, Huang2013b, 
    Oh2017}), optical colour gradients (e.g., \citealt{LaBarbera2010, LaBarbera2012}), 
    and stellar population gradients \citep[e.g.,][]{Coccato2010, Coccato2011, 
    Greene2015, Barbosa2016}. 
    These observations are generally consistent with the two-phase formation scenario. 
    However, it is still not clear whether this picture correctly 
    predicts the connection between the stellar mass distributions in massive galaxies 
    and their dark matter haloes.
    
    In the $\Lambda$CDM cosmology, the assembly of massive ETGs is intrinsically 
    tied to the hierarchical growth of their host dark matter haloes 
    (e.g., \citealt{Leauthaud2012, Behroozi2013, Shankar2013}). 
    Hydrodynamic simulations suggest that the fraction of stars accreted through 
    mergers (the \textit{ex situ} component) in central galaxies increases with halo 
    mass (e.g., \citealt{RodriguezGomez2016, Pillepich2017b}). 
    The major merger rate is not a strong function of progenitor halo mass 
    (e.g., \citealt{Shankar2015}) but minor mergers rate should increase with 
    halo mass, hence play an important role in determining the structures of central 
    galaxies (e.g., \citealt{Guo2011, Yoon2017}). 
    Minor mergers are efficient at `puffing up' the outskirts of massive 
    galaxies (e.g., \citealt{Oogi2013, Bedorf2013}). 
    Because the minor merger rate increases with halo mass, the 
    structures of massive ETGs and the well-known stellar mass--effective radius 
    relation (\mstar{}--$R_{\rm e}$; e.g., \citealt{Shen2003, Guo2009}) should 
    depend on their `environment'\footnote{There are multiple definitions of 
    `environment' in the literature. 
    In this work, we use `environment' and halo mass interchangeably.}
    (e.g., \citealt{Shankar2013, Shankar2014}). 
    However, evidence for the environment-dependence of \mstar{}--$R_{\rm e}$
    at low redshift is still not very solid (\citealt{Nair2010, HCompany13}; 
    but also see \citealt{Yoon2017}), and the results at higher redshift are 
    even more unclear (e.g., \citealt{Papovich2012, Lani2013, Delaye2014}; but 
    also see \citealt{Rettura2010}).
	
    Deep images of massive galaxies can probe their outer stellar halos of massive 
    galaxies and test these predictions. 
    Unfortunately, this is observationally challenging since the stellar haloes 
    of massive galaxies can extend to $>100$ kpc (e.g., 
    \citealt{Tal2011, DSouza2014}), and their surface brightness profiles decline 
    rapidly with typical values of $\mu > 26.0$ \sb{} in $i$-band at 100 kpc and 
    at $z\sim0.3$. 
    In \citet[][Paper I hereafter]{hscMassiveI}, we showed that deep, 
    multi-band imaging from the Subaru Strategic Program (SSP; \citealt{HSC-SSP,
    HSC-DR1}) using Hyper Suprime-Cam (HSC; \citealt{Miyazaki2012}, 
    Miyazaki in~prep.) allows us to extract robust surface stellar
    mass density (\mden{}) profiles for {\it individual} galaxies with 
    \logms{}$>11.4$ at $0.3 < z < 0.5$ and out to 100 kpc. 
    In Paper I, we characterized the stellar mass profiles of massive ETGs and 
    showed that there is a large intrinsic scatter in the stellar haloes of
    massive galaxies on 100-kpc scales. 
    In this paper, we investigate whether the large scatter in the outer 
    profiles of massive galaxies correlates with halo mass. 
    We conclusively show that the sizes and stellar haloes of massive central 
    galaxies depend on dark matter halo mass. 
    In other words, we reveal the halo mass dependence of the mass--size relation 
    for massive ETGs.   
    
    This paper is organized as follows. 
    In \S \ref{sec:data} we briefly introduce the sample selection and the data 
    reduction processes.  
    Please refer to \citet{hscMassiveI} for more technical details.
    Our main results are presented in \S \ref{sec:result} and discussed in 
    \S \ref{sec:discussion}. 
    Our summary and conclusions are presented in \S \ref{sec:summary}.

    Magnitudes use the AB system (\citealt{Oke1983}), and are corrected for galactic 
    extinction using calibrations from \citet{Schlafly11}. 
    We assume $H_0$ = 70~km~s$^{-1}$ Mpc$^{-1}$, ${\Omega}_{\rm m}=0.3$, 
    and ${\Omega}_{\rm \Lambda}=0.7$.
    Stellar mass is denoted \mstar{} and has been derived using a Chabrier initial mass 
    function (IMF; \citealt{Chabrier2003}). Halo mass is defined as 
    $M_{\rm 200b}\equiv M(<r_{\rm 200b})=200\bar{\rho} 
    \frac{4}{3}\pi r_{\rm 200b}^3$, where $r_{\rm 200b}$
    is the radius at which the mean interior density is equal to 200 times
    the mean matter density ($\bar{\rho}$). 
    As in \citet{hscMassiveI}, we do not attempt to decompose or distinguish any  
    potential `intra-cluster' 
    component (ICL; e.g., \citealt{Carlberg1997, Lin2004, Gonzalez2005, Mihos2005}). 
    

  \begin{figure*}
      \centering 
      \includegraphics[width=\textwidth]{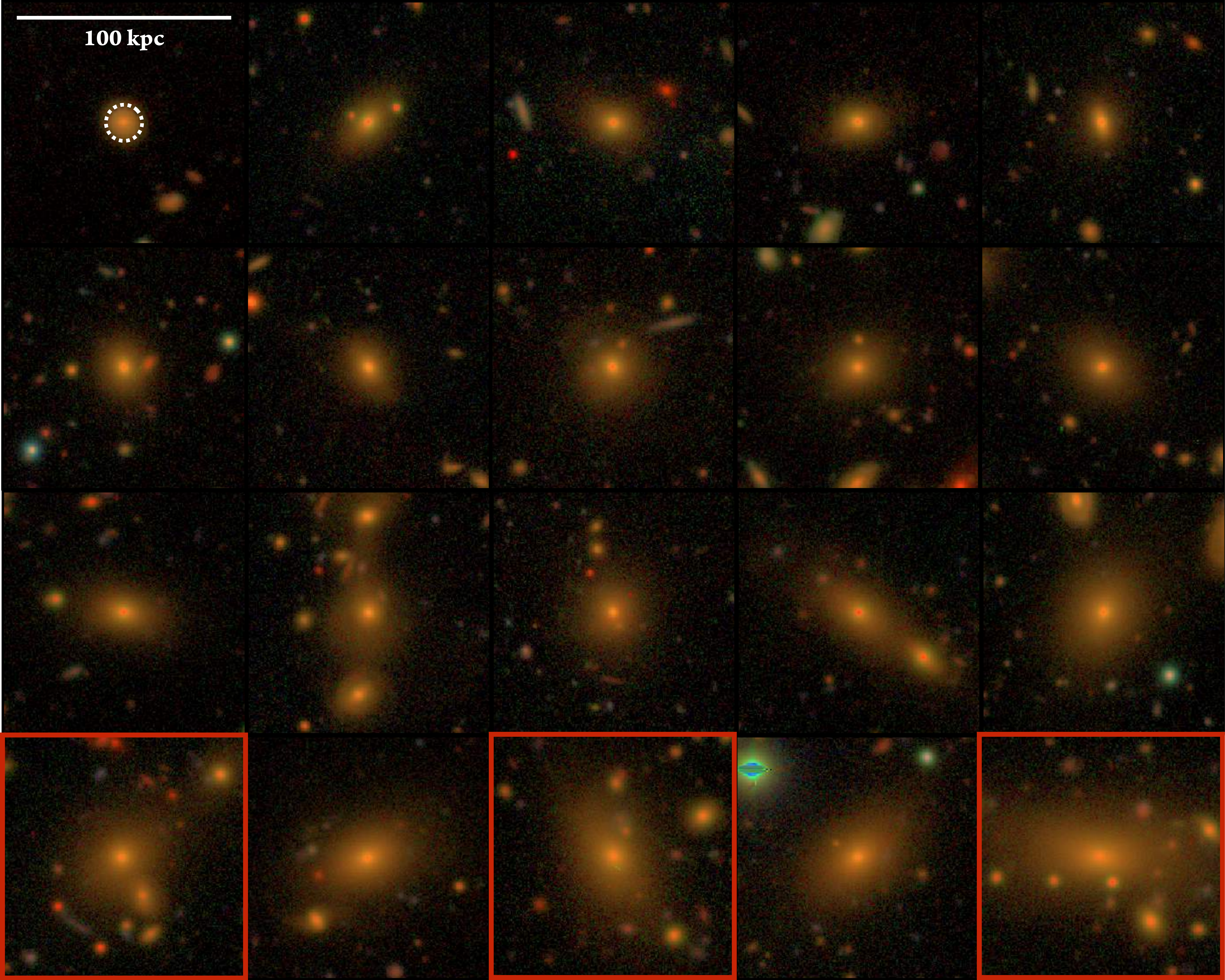}
      \caption{
          Three-colour images for a subsample of massive galaxies at $z{\sim}0.4$. 
          All of these massive galaxies have very similar \mstar{} within a 10-kpc 
          elliptical aperture 
          ($11.2<\log_{10} (M_{\star,10\ \mathrm{kpc}}/M_{\odot})<11.3$). 
          The dashed-line circle at the top-left figure indicates $R=$10 kpc.
          These galaxies are rank-ordered from top to bottom and from left to right 
          by their \mstar{} within a 100-kpc elliptical aperture that varies 
          from $10^{11.2}\ M_{\odot}$ to $10^{11.7}\ M_{\odot}$. 
          At fixed `inner' mass (\minn{}), massive galaxies display significant
          diversity in their outer profiles. 
          Red boxes indicate galaxies from dark matter haloes that are more massive 
          than ${\sim} 10^{14}\ M_{\odot}$. 
          }
      \label{fig:m100_m10_color}
  \end{figure*}

\section{Sample Selection and Data Reduction}
    \label{sec:data}
    
    We refer the reader to Paper I for an in-depth description of the sample selection 
    and data reduction processes. 
    Here, we briefly summarize the main steps.
    
    We use imaging data from the HSC internal data release 
    \texttt{S15B}, which is very similar to the Public Data Release 1 
    (\citealt{HSC-DR1} and covers ${\sim} 110$ deg$^2$ in all five-band ($grizy$) to 
    the full depth in the wide field. 
    The data are reduced by \texttt{hscPipe 4.0.2}, a derivative of the 
    Large Synoptic Survey Telescope (LSST) pipeline (e.g.\ \citealt{Juric2015}; 
    \citealt{Axelrod2010}), modified for HSC (\citealt{HSC-PIPE}).
    The pixel scale of the reduced image is $0.168$\asec{}.
    We use $i$-band images for extracting surface brightness profiles. 
    HSC $i$-band images are typically 3--4 mag deeper than SDSS 
    (Sloan Digital Sky Survey; e.g., \citealt{SDSS-DR7, SDSS-DR8, SDSS-DR12})  
    and have superb seeing conditions (mean $i$-band seeing has FWHM$=0.6$\asec{}).
    
    In Paper~I, we select a sample of 25286 bright galaxies with spectroscopic 
    redshifts or reliable `red-sequence' photometric redshifts (\citealt{Rykoff2014}) 
    at $0.3<z<0.5$. 
    Within this redshift range, we have a large enough volume 
    ($\sim5\times 10^6$ Mpc$^3$) to sample the galaxy stellar mass function above 
    \logms$>11.6$, and we can spatially resolve galaxies profiles to $\sim 5$ kpc 
    ($1.0^{\arcsec}$ corresponds to 4.4 and 6.1 kpc at $z=0.3$ and 0.5, respectively). 
    Massive galaxies should experience little structural evolution and 
    size growth between $z=0.5$ and 0.3 ($\sim$1.5 Gyr time span) 
    based on model predictions (e.g., \citealt{Shankar2015}).
    
    After carefully 
    masking out surrounding neighbors and accounting for the subtraction of the 
    background light, we derive $i$-band surface brightness profiles out to 100 kpc. 
    We use the broadband spectral energy distributions (SED) fitting code 
    \texttt{iSEDFit}\footnote{http://www.sos.siena.edu/~jmoustakas/isedfit/} 
    (\citealt{Moustakas13}) to measure \m2l{} ratios and $k$--corrections using 
    five-band forced \cmodel{} magnitudes from \texttt{hscPipe}. 
    We assume a \citet{Chabrier2003} IMF, the Flexible Stellar Population 
    Synthesis models\footnote{http://scholar.harvard.edu/cconroy/sps-models}
    (FSPS; \texttt{v2.4}; \citealt{FSPS}, \citealt{Conroy2010}), the 
    \citet{Calzetti2000} extinction law, and a simple delayed-$\tau$ model for 
    star formation histories (SFH). 
    Using HSC data, we can measure the \mden{} profiles of massive galaxies to 
    $\sim 100$ kpc, and we integrate these profiles within elliptical isophotal
    apertures at different physical radii. 
    As explained in Paper~I, we focus on the two following metric masses:
        
    \begin{itemize}
    
        \item Stellar mass within 10 kpc (hereafter noted \minn{}), which we use 
            as a proxy for the stellar mass of the \textit{in situ} stellar 
            component. 
            This is motivated both by observations and simulations 
            (e.g.~\citealt{vanDokkum2010}, \citealt{RodriguezGomez2016}). 
            The value of 10 kpc that we quote here corresponds to the radius of the 
            major axis of the isophotal ellipse.
            
        \item Stellar mass within 100 kpc (hereafter noted \mtot{}). 
            We use \mtot{} as a proxy for the `total' stellar mass. 
            In Paper~I we show that \mtot{} recovers more light compared to 
            HSC \cmodel{} photometry with differences that can be a large as 0.2 dex 
            in magnitude.        
               
   \end{itemize}
   
   We use these two simple metric masses to explore the \mstar{}-dependence of the 
   fraction of accreted stars and to reveal the diversity of stellar envelopes among 
   massive galaxies. 
   In practice, we also have the full profiles for each galaxy and can cast our 
   results in terms of the full stellar mass profiles. 
   However, in many cases we find it useful to display figures using the simpler 
   \minn{} and \mtot{} quantities. 
   In Figure \ref{fig:m100_m10_color}, we highlight the diversity of galaxies as a 
   function of \minn{} and \mtot{}. 
   Figure \ref{fig:m100_m10_color} shows a subsample of massive galaxies with very 
   similar \minn{} but show a large range of \mtot{}. 
   We use these two aperture masses to guide our comparison of massive galaxies 
   as a function of environment.

  \begin{figure*}
      \centering 
      \includegraphics[width=\textwidth]{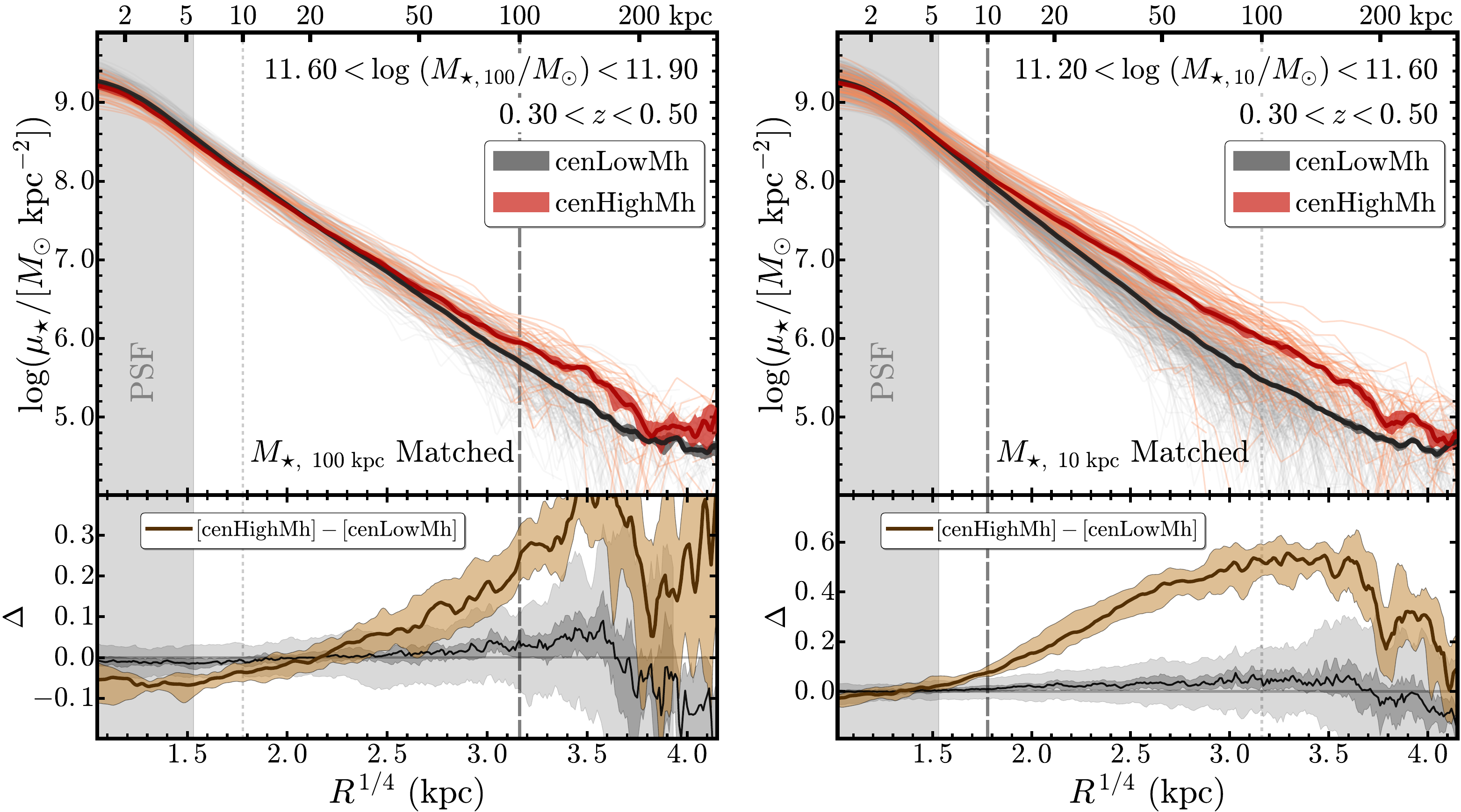}
      \caption{
          The environmental dependence of the stellar mass profiles of massive central 
          galaxies. 
          \textbf{Left}: Halo mass dependence of galaxy \mden{} profiles at fixed 
          `total' stellar mass (\mtot{}). 
          \textbf{Right}: Halo mass dependence of galaxy \mden{} profiles at fixed 
          `inner-mass' (\minn{}). 
          In each plot, we match the two samples so that they share similar 
          distributions of mass and redshift (Appendix~\ref{app:match}).
          We use $R^{1/4}$ as $x$--axis to provide a balanced view of both the central 
          and outer regions of the galaxy.
          Orange and red lines correspond to central galaxies living in 
          haloes with \logmh{} $\geq 14.2$. 
          Black and grey lines correspond to central galaxies living in haloes with 
          \logmh{} $\leq 14.0$. 
          Thin lines show the profiles of individual galaxies, while thick lines show 
          the median profile. 
          The uncertainty on the median profile is given by the shaded region and is 
          computed via bootstrap resampling. 
          Brown lines in the bottom panels show the relative difference between  
          the two median profiles 
          ($\Delta = \log(\mu_{\star, \mathrm{cenHighMh}}) - 
          \log(\mu_{\star, \mathrm{cenLowMh}}$). 
          Errors in the difference between the two profiles are also computed 
          via bootstrap.
          The grey-shaded regions show a Monte Carlo test to assess how likely it is to 
          obtain $\Delta$ from random sub-samples of the data. 
          To compute the grey-shaded regions, we first mix the two samples 
          (\rbcg{} and \nbcg{}), then draw sub-samples of galaxies from the mixed 
          population and compute $\Delta$ in the same fashion as for our fiducial 
          signal. 
          We repeat this process 5000 times.  
          The dark grey--shaded region (light grey--shaded region) shows the 1-$\sigma$ 
          (3-$\sigma$) fluctuations in $\Delta$ from these 5000 draws.
          }
      \label{fig:prof_1} 
  \end{figure*}

  \begin{figure*}
      \centering 
      \includegraphics[width=\textwidth]{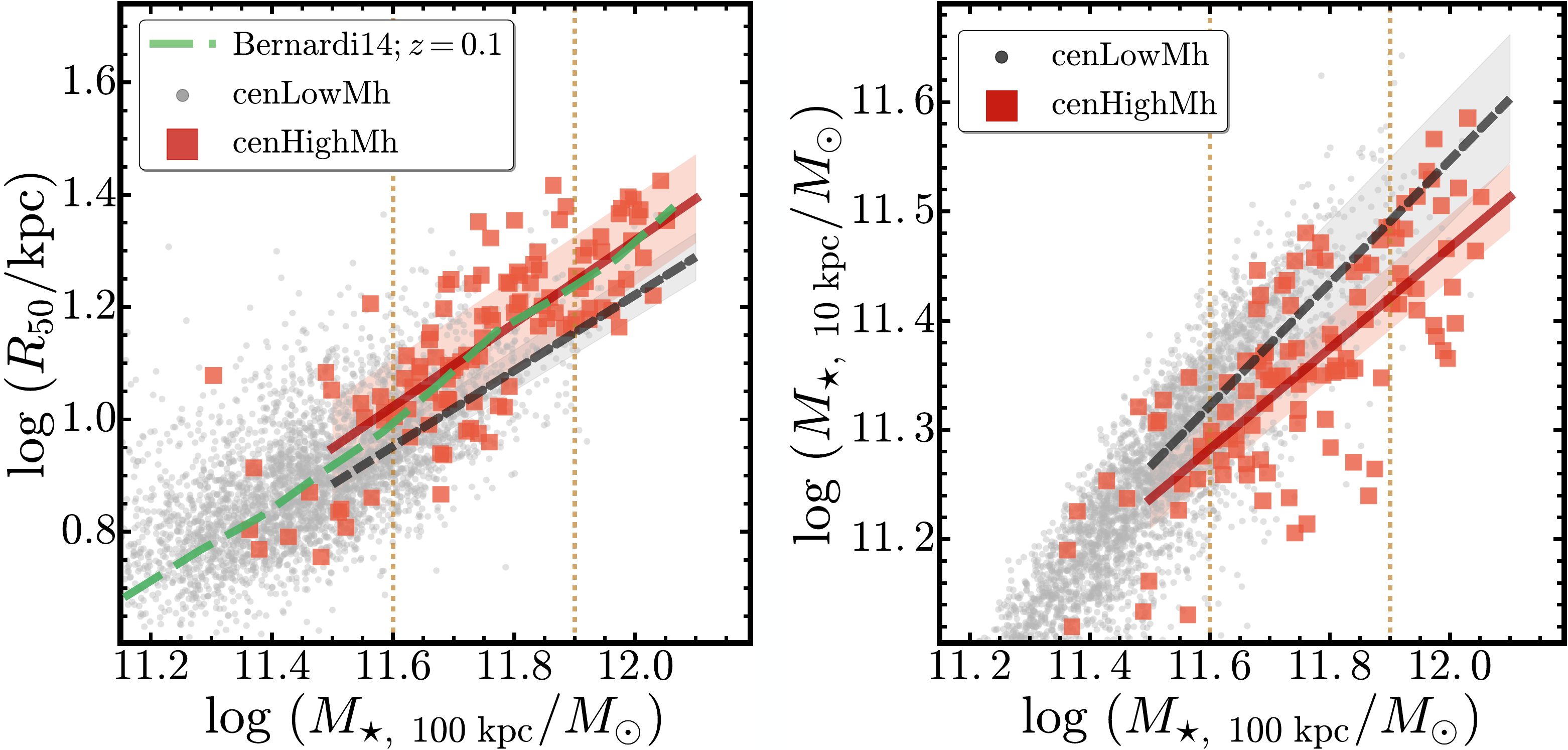}
      \caption{
          \textbf{Left}: The mass-size relations for \rbcg{} (orange squares) and 
          \nbcg{} (grey dots) galaxies. 
          Two vertical lines highlight our $11.6<$ \logms{} $<11.9$ mass bin. 
          The red solid line shows the best-fitting mass--size relation for \rbcg{} and the 
          grey dashed line shows the best-fitting relation for \nbcg{}. 
          Shaded regions in lighter colours show the 1-$\sigma$ uncertainties
          from Markov chain Monte Carlo (MCMC) sampling.  
          \song{
          The green dashed line shows the mass--size relation for $z\sim 0.1$ 
          early--type galaxies from \citet{Bernardi2014}. 
          }
          \textbf{Right:} The relations between \mtot{} and \minn{} for the 
          \rbcg{} and \nbcg{} galaxies, along with the best-fitting scaling relations for 
          both samples.
          }
      \label{fig:scaling_relation} 
  \end{figure*}

\subsection{Massive Central Galaxies from Different Environments}
    \label{ssec:cen}
         
    In this work, we focus on massive galaxies with \mtot{} $>10^{11.6}$~\msun{}. 
    In Paper~I, we demonstrate that this sample is almost mass complete over our full 
    redshift range. 
    In addition to the mass cut above, we also limit our sample to galaxies that live 
    at the centers of their own dark matter haloes -- so-called `central' galaxies. 
    We use the \redm{} \texttt{v5.10} \citep{Rykoff2014, Rozo2015b} cluster catalogue 
    to help us construct two central galaxy samples: one for high mass haloes, and 
    one for low mass haloes.
    
 
    First, we build a sample of central galaxies in high mass haloes. 
    We select 68 massive central galaxies from \redm{} clusters with richness
    $\lambda \geq 30$, with central probability $P_{\mathrm{Cen}} \geq 0.7$, and 
    \logmtot{}$ >11.5$ at $0.3 < z < 0.5$ (63/68 have \logmtot{}$ >11.6$). 
    This $\lambda$ limit is chosen to mitigate incompleteness in the cluster catalogue
    at the high end of our redshift window. 
    The $P_{\mathrm{Cen}}$ limit is imposed to limit our sample to central galaxies. 
    \citet{Simet2017} present a calibration of the \mhalo{}-$\lambda$ relation for 
    \redm{} clusters using SDSS weak-lensing. 
    Based on this calibration, our sample corresponds to central galaxies living 
    in haloes with \mhalo{}$>10^{14.2}$~\msun{}. 
    This calibration is consistent with several other independent calibrations 
    using different methods (e.g., \citealt{Saro2015, Farahi2016, 
    Melchior2016, Murata2017}). 
    The median richness of the sample is $\lambda \approx 41$ 
    (\logmhalo{}$\approx 10^{14.3}$), and there are 44 central galaxies 
    in clusters with $\lambda>50$ (\mhalo{}$\approx 10^{14.5}$).
    We refer to this sample of \textbf{central galaxies in massive haloes} as 
    the \rbcg{} sample.
    
    Second, we build a sample of central galaxies in low mass haloes. 
    We begin by excluding all galaxies in \redm{} clusters with $\lambda > 20$.
    We convert $\lambda$ to $M_{\mathrm{200b}}$ using the \citet{Simet2017} calibration. 
    For each cluster, we compute $R_{\mathrm{200b}}$ using the \texttt{Colossus} Python 
    package 
    (\citealt{Colossus})\footnote{\url{http://www.benediktdiemer.com/code/colossus/}}
    provided by \citet{Diemer2015}
    We exclude all galaxies within a cylinder around each cluster, with a radius
    equal to $R_{\mathrm{200b}}$, and a length equal to twice the value of the 
    photometric redshift uncertainty of the cluster (typically around 0.015 to 0.025).
    This second sample is dominated by central galaxies living in haloes with
    $M_{\mathrm{200b}} < 10^{14}$~\msun{}; we refer to this sample as \nbcg{}. \
    There are 833 central galaxies with \logmtot{}$> 11.6$ in this sample.

    Given the high stellar mass, satellite contamination in our sample should be 
    low (e.g., \citealt{Reid2014, Hoshino2015, Saito2016, vanUitert2016}). 
    For instance, the model from \citet{Saito2016} predicts that our \rbcg{} sample 
    should only contain $\sim 7$\% satellites (corresponding to satellites with 
    \logmtot{}$>11.6$ and living in haloes with \logmh$<14.0$).
    
    Appendix~\ref{app:basic} shows the distributions of redshift, \mtot{}, and 
    \minn{} for the two samples. 
    We also compare these two samples on a \mtot{} versus rest--frame $(g-r)$ colour 
    plane. 
    Both samples follow the same red-sequence, with only a handful of galaxies 
    displaying bluer colours.
    Given the available calibration, the current $\lambda$ cut should ensure the 
    \rbcg{} and \nbcg{} samples have significant difference in average halo mass, 
    although we can not directly estimate the average \logmh{} for the \nbcg{}
    sample. 
    In Appendix~\ref{app:robust}, we show that the main results are robust even when 
    $\lambda > 20$ cut is adopted for the \rbcg{} sample.
        
    Our analysis fails to extract 1-D profiles for $\sim10$\% of 
    \rbcg{} and \nbcg{} galaxies due to ongoing major mergers or projection effects 
    (e.g.\ nearby foreground galaxy or bright stars). 
    We exclude these galaxies from our analysis and this low failure rate should not 
    affect any of our results.
    

\section{Results}
    \label{sec:result}
    
    As shown in Figure~\ref{fig:m100_m10_color}, massive central galaxies at fixed  
    \minn{} display a large diversity in their stellar haloes. 
    In Paper~I, we explored the \mstar{}-dependence of these stellar haloes. 
    We now investigate the relation between \mden{} profiles, stellar haloes, and 
    dark matter halo mass. 
    We remind the reader that although a circular aperture is shown on 
    Fig~\ref{fig:m100_m10_color}, in practice we extract 1-D \mden{} profiles and 
    estimate \mtot{} and \minn{} using elliptical apertures following the average 
    flux-weighted isophotal shape. 

\subsection{Environmental Dependence of the Stellar Mass Density Profiles of Massive 
            Galaxies}
    \label{ssec:sbp_mtot} 
       
    First, we ask whether the \mden{} profiles of massive central galaxies depend on 
    halo mass at fixed stellar mass.    
    We show comparisons of \mden{} profiles at both fixed \mtot{} and fixed 
    \minn{} (see Figure~\ref{fig:prof_1}). 
    All comparisons are performed with a fixed underlying redshift
    distribution by matching samples in redshift in addition to stellar mass
    (see Appendix~\ref{app:match}, Appendix~\ref{app:redshift}, 
    and Fig~\ref{fig:match} for details). 
   
    Fig~\ref{fig:prof_1} compares the \mden{} profiles of massive central galaxies in 
    low mass haloes to those in high mass haloes at fixed \mtot{} (left panel) and at 
    fixed \minn{} (right panel). 
    The left panel compares galaxies that have similar `total' stellar mass. 
    The right panel uses \minn{} as a proxy for the \textit{in situ} component to 
    compare the profiles of galaxies that presumably have similar early formation 
    histories, but which live in different dark matter haloes today.
    This figure shows the main result of this paper, namely that \emph{the \mden{} 
    profiles of massive central galaxies show a clear dependence on dark matter halo 
    mass at both fixed \mtot{} and \minn{}.}

    We estimate the uncertainties of the median \mden{} profiles using a bootstrap 
    resampling test, and we perform statistical tests to demonstrate that the 
    difference between the profiles is more significant than the level allowed by 
    the intrinsic randomness within the combined \rbcg{} and \nbcg{} sample. 
    We also conduct a variety of tests that verify the robustness of these results with
    respect to our \mtot{} binning scheme, $\lambda$ cut, the redshift range, and the 
    choice of apertures used for the metric masses. 
    Please see Appendix~\ref{app:robust} for further details.
    Appendix \ref{app:cog} and Figure~\ref{fig:cog} compares the same 
    \mtot{}-matched samples using cumulative \mstar{} profiles (`curve of growth') 
    and the fraction of \mtot{} enclosed within different radii.
    Both comparisons highlight the differences in the median \mden{} profiles 
    from different angles.   
   
    The key features in Figure~\ref{fig:prof_1} are the following:
    
    \begin{itemize}
        
        \item At fixed \mtot{}, central galaxies in high mass haloes display shallower 
            \mden{} profiles compared to those in low mass haloes (i.e., they have 
            flatter inner \mden{} profiles and more significant outer stellar 
            envelopes). 
            
        \item The median \mden{} profiles of the two samples cross 
            each other at ${\sim} 15$-20 kpc, roughly the typical effective radius 
            ($R_{\mathrm{e}}$) of galaxies at these masses (\logmtot{}$>11.7$--11.8).  
                        
        \item When matched by \mtot{}, differences in the inner regions appear 
            to be small, but this is also driven by the use of a logarithmic y-axis. 
            The difference becomes more apparent at $R>50$ kpc.  
                                
        \item Massive galaxies matched by \minn{} display a range of \mtot{} values. 
            Those in massive dark matter halos have more prominent outer stellar haloes. 
            The scatter in the outer profiles observed in Figure \ref{fig:prof_1} is 
            an \textit{intrinsic} scatter (not measurement error).     
                            
    \end{itemize}

    Fig \ref{fig:prof_1} shows that the environmental dependence of the profiles of 
    massive central galaxies is a subtle effect that is most prominent at large radii 
    ($R>50$ kpc). 
    This may explain why previous attempts to detect this effect using shallower images 
    have often failed. 
    
    The effect becomes more pronounced for even more massive galaxies. 
    This is shown in Fig~\ref{fig:prof_2} in Appendix~\ref{app:robust}.
    
    In summary, we detect a subtle, but robust halo mass dependence of the profiles of 
    massive central galaxies. 
    This dependence could be driven by the fact that massive halos have a larger minor 
    merger rate compared to less massive haloes. 
    Non--dissipative (minor) mergers should not strongly alter inner profiles, but can
    efficiently build up outer haloes (e.g., \citealt{Hilz2013}, \citealt{Oogi2013}).
      

\subsection{The Environmental Dependence of Scaling Relations}
    \label{ssec:scaling}
    
    We have shown that the \mden{} profiles of massive galaxies vary with the masses 
    of their host dark matter haloes. 
    We now turn our attention to the more commonly studied stellar mass--size relation
    (\mstar{}--$R_{\mathrm{e}}$). 
    In addition, we consider halo mass dependence on the \mtot{}--\minn{} plane. 
    
\subsubsection{Mass--Size Relation}
    \label{sssec:mass_size}
    
    The tight relation between \mstar{} and effective radius (or half-light radius; 
    $R_{\mathrm{e}}$ or $R_{\mathrm{50}}$; e.g., \citealt{Shankar2013, Leja2013, 
    vdWel2014}) is one of the most important scaling relationships for ETGs. 
    Despite numerous attempts, previous studies have failed to detect the 
    \mhalo{}-dependence of the \mstar{}--$R_{\mathrm{e}}$ relation at low-$z$ 
    (e.g., \citealt{Weinmann2009, Nair2010, HCompany13, Cerbrian2014}; 
    except for the recent result by \citealt{Yoon2017}). 
    
    However, `size' is not a well-defined parameter for massive galaxies with very 
    extended stellar mass distributions. 
    In practice, measurements of the `effective radius', or `half-light radius', depend 
    on resolution, depth, filter, and may also depend on the adopted model for the 
    light profile.  
    This makes comparisons of size measurements among different observations, 
    or between observations and models, uncertain. 
    This is the main reason why we prefer to compare \mden{} profiles directly which 
    completely bypasses the need to consider `size'.
    
    Nonetheless, to enable comparisons with past work, we now consider the more 
    traditional mass-size relation. 
    We adopt the radius enclosing 50\% of stellar mass within 100 kpc 
    ($R_{\mathrm{50}}$; derived from the $i$-band curve-of-growth) as our `size' for 
    massive galaxies. 
    This definition of `half-light radius' is more robust against structural details, 
    model choice, and background subtraction, compared to the effective radius measured 
    using oversimplified 2-D models such as the single-\ser{} model. 
    Massive galaxies in this sample are large enough so that the impact of seeing 
    is not a concern.
    
    The left panel of Fig~\ref{fig:scaling_relation} shows the mass-size relation for 
    our two samples. 
    We fit the \logmtot{}-$\log_{10} (R_{\mathrm{50}}/\mathrm{kpc})$ relations at 
    $\log(M_{\star,100})>11.6$ using the \texttt{emcee} MCMC sampler 
    (\citealt{Emcee})\footnote{The initial guesses are based on maximum 
    likelihood estimates, and we assume flat priors for parameters.}.
    Uncertainties in both \mtot{} and $R_{\mathrm{50}}$ are considered. 
    For \logmtot{}$>11.6$ galaxies, the typical uncertainty for mass is $\sim 0.12$
    dex.  
    The $R_{\mathrm{50}}$ uncertainty is based on the \mden{} profile and is very small 
    due to the high \s2n{} of the profile. 
    We manually assign a 10\% error for $R_{\mathrm{50}}$. 
    
    The best-fitting relation for \rbcg{} is:
    
    \begin{equation}
        \begin{aligned}
        \log_{10} (R_{\mathrm{50}}/\mathrm{kpc}) = & (0.74\pm0.13) \times \log_{10} (M_{\star, 100\ \mathrm{kpc}}/M_{\odot}) \\ & -(7.56\pm1.56)
        \end{aligned}
    \end{equation}

    \noindent And for \nbcg{}, we find:
    
    \begin{equation}
        \begin{aligned}
        \log_{10} (R_{\mathrm{50}}/\mathrm{kpc}) = & (0.68\pm0.06) \times \log_{10} (M_{\star, 100\ \mathrm{kpc}}/M_{\odot}) \\ & -(6.88\pm0.75)
        \end{aligned}
    \end{equation}
    
    \noindent As shown in the left panel of Fig~\ref{fig:scaling_relation}, the two 
    samples lie on \mtot{}--$R_{\mathrm{50}}$ relations that have similar slopes but 
    different normalizations. The best-fitting mass--size relation derived here suggests 
    that, at $0.3 < z < 0.5$, central galaxies with \logmtot{}$>11.6$ in \logmhalo{}$>14.2$
    haloes are on average $\sim20$\% larger than centrals in \logmhalo{}$<14.0$
    haloes at fixed \mtot{}.
    This result is robust against the stellar mass range over which the fit is 
    performed and against the definitions of  `total' \mstar{} and half-light 
    radius\footnote{Using \mstar{} within 120 or 150 kpc, or using the 
    $R_{\mathrm{50}}$ derived within these apertures does not change the results.}. 
    
    Due to the steep slope of the mass--size relation, secondary binning 
    (e.g., different halo mass) may introduce artificial differences 
    (e.g., \citealt{Sonnenfeld2017}).
    Besides the differences in best--fit \mtot{}--$R_{\mathrm{50}}$ relation, the 
    median $R_{\mathrm{50}}$ of the \mtot{}-matched samples also confirm the above 
    conclusion. 
    In addition, we use the normalized size parameter ($\gamma$; e.g., 
    \citealt{Newman2012, HCompany13}) to further test our results. 
    In \citet{HCompany13}, $\gamma$ is defined as:
    
    \begin{equation}
        \log_{10}(\gamma) = \log_{10} (R_{\mathrm{50}}) + \beta (11 - \log_{10}M_{\star, 100\ \mathrm{kpc}}),
    \end{equation}
    
    where $\beta$ is the slope of the mass--size relation.
    We estimate the average $\gamma$ of both samples at \logmtot{}$>11.6$. 
    For \rbcg{}, $<\gamma> = 4.2\pm0.4$ and for \nbcg{}, 
    $<\gamma> = 3.8\pm0.3$. 
    The environmental dependence of $R_{\mathrm{50}}$ at fixed \mtot{} is more 
    significant than the \citet{HCompany13} result but weaker than some 
    model predictions (e.g., \citealt{Shankar2014}).
    
    
    We also compare with the mass--size relation from the Figure~12 of 
    \citealt{Bernardi2014} (green dashed line).
    These authors studied the mass--size relation for a large sample of $z\leq 0.1$ 
    ETGs by fitting their SDSS images with a 2--component model that 
    consists of a \ser{} and an exponential component (\texttt{SerExp}). 
    Comparing to the single-\ser{} model, the \texttt{SerExp} model provides 
    much less biased measurements of total luminosity and effective radius 
    for massive galaxies.  
    The stellar masses are derived based on a \m2l{}--color relation for 
    SDSS $r$-band assuming a Chabrier IMF (see \citealt{Bernardi2010}).
    The mass-size relation from \citealt{Bernardi2014} is qualitatively 
    consistent with the one from this work.   
    Differences of redshift and assumptions in stellar mass measurements between 
    \citealt{Bernardi2014} and this work can lead to systematic shifts on
    the mass--size plane. 
    However, it is still interesting that the mass-size relation derived by 
    2--component model fitting on much shallower SDSS images has very similar 
    slope comparing to the HSC result. 
    The impacts of imaging depth and modeling method on the study of mass--size 
    relation deserves further investigation using a common sample of massive 
    galaxies in the near future. 
    
    
\subsubsection{\mtot{} - \minn{} Relation}
    \label{sssec:m100_m10}
    
    We now explore the environment dependence of galaxy structure using the 
    \mtot{}-\minn{} relation.
    Compared to the mass--size relation, the \mtot{}--\minn{} relation is not plagued 
    by the ambiguity of galaxy `size', and it also enables a more straightforward 
    comparison with numerical simulations.
    
    The right panel of Fig \ref{fig:scaling_relation} compares our two samples on the 
    \mtot{}--\minn{} plane. 
    The two samples follow distinct best-fitting \mtot{}--\minn{} relations. 
    For \rbcg{} galaxies with \logmtot{}$>11.6$ we find:
    
    \begin{equation}
        \begin{aligned}
        \log_{10} (M_{\star, 10\ \mathrm{kpc}}/M_{\odot}) = & (0.48\pm0.06) \times \log_{10} (M_{\star, 100\ \mathrm{kpc}}/M_{\odot}) \\ & +(5.72\pm0.75).
        \end{aligned}
    \end{equation}
    
    \noindent In the same range of \mtot{}, the best-fitting relation for \nbcg{} is:
     
    \begin{equation}
        \begin{aligned}
        \log_{10} (M_{\star, 10\ \mathrm{kpc}}/M_{\odot}) = & (0.56\pm0.03) \times \log_{10} (M_{\star, 100\ \mathrm{kpc}}/M_{\odot}) \\ & +(4.82\pm0.30).
        \end{aligned}
    \end{equation}
    
    These results are robust against the exact choice of the stellar mass range over 
    which the fit is performed.
    These results are also unchanged when we replace \minn{} with the stellar 
    mass within a 5- or 15-kpc aperture, or if \mtot{} is replaced with a stellar mass 
    within a 120- or 150-kpc aperture.  
    
    Figure \ref{fig:scaling_relation} presents the same conclusions as in the previous 
    section, namely that at fixed \mtot{}, central galaxies of more massive haloes tend 
    to have a smaller fraction of stellar mass in their inner regions and more prominent 
    outer stellar haloes. 
    And at fixed \minn{}, central galaxies of more massive haloes on average are 
    $\sim 0.1$ dex more massive than the ones from less massive haloes within a 100 kpc 
    aperture, which corresponds to $\sim 10^{11} M_{\odot}$ of stellar mass differences. 
    If we can assume that the same \minn{} suggests similar \mstar{} when they were just
    quenched at high redshift (or similar \textit{in situ} stellar mass), this means 
    the central galaxies of \logmh{}$>14.2$ haloes typically experienced one more major 
    merger or a few more minor mergers comparing to the ones of \logmh{}$<14.0$ haloes.
    It would be interesting to compare this prediction with hydro-simulations or 
    semi-analytic models. 
      

  \begin{figure*}
      \centering 
      \includegraphics[width=\textwidth]{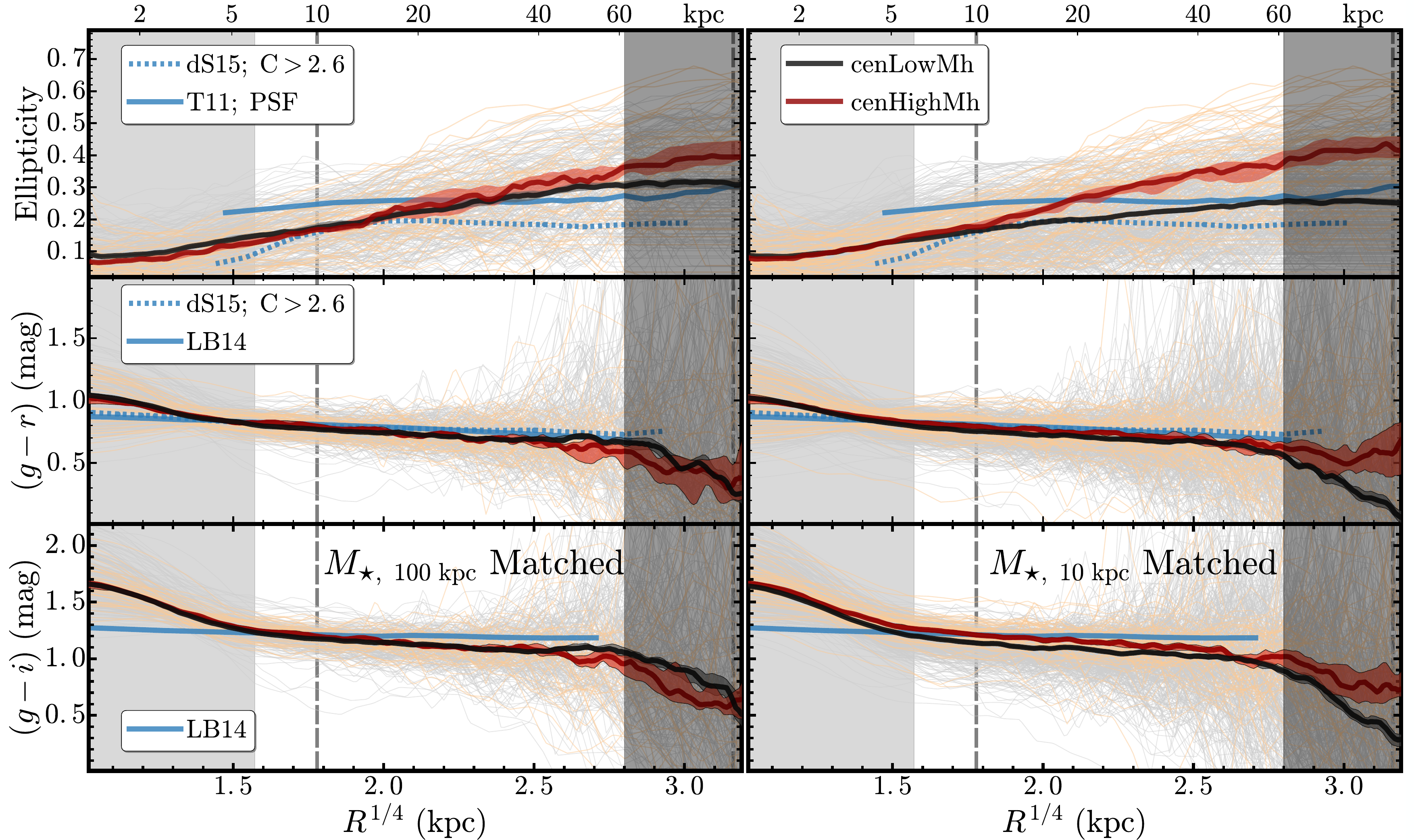}
      \caption{
          Radial variations in ellipticity and optical colours for massive galaxies. 
          The format of this figure is similar to the right hand side of 
          Fig~\ref{fig:prof_1}. 
          Top panels show the ellipticity profile, middle panels show $g-r$, and lower 
          panels show $g-i$. 
          We compare our results with those from \citet{Tal2011}
          (solid blue line on top panels). 
          We also compare our results with those from a stacking analysis of nearby 
          massive galaxies with high concentration indices ($C>2.6$) from 
          \citet{DSouza2014} (blue dashed lines in top and middle panels). 
          We also compare our results with the average $g-r$ and $g-i$ colour profiles 
          from a large sample of nearby elliptical galaxies for \citet{LaBarbera2010}
          (blue, solid lines in middle and bottom panels).
          }
      \label{fig:ell_color}
  \end{figure*}

\subsection{Ellipticity and Colour Profiles}
    \label{ssec:ell_color}
    
    In Paper~I, we show that the ellipticity of the outer stellar halo increases with 
    stellar mass but that rest-frame colour gradients do not depend strongly on 
    stellar mass. 
    In this paper, we take this analysis one step further to investigate whether either 
    of these quantities depends on halo mass. 
    We focus on ellipticity and colour profiles within 5--60 kpc where we can ignore
    differences in sky subtraction and seeing across different filters. 
    Galactic extinction and $k$ corrections are applied to both $(g-r)$ and $(g-i)$ 
    colour profiles.
    
    Figure \ref{fig:ell_color} shows the average ellipticity, $g-r$, and $g-i$ colour 
    profiles for galaxies at fixed \mtot{} and fixed \minn{}. Our main findings are: 
    
    \begin{itemize}
    
        \item The ellipticity profiles of massive central galaxies do not depend on 
            halo mass at fixed \mtot{} at $R<60$ kpc (upper left panel).     
                
        \item However, we do find that at fixed \minn{}, galaxies in massive halos 
            have more elliptical outer stellar halos compared to galaxies in low mass 
            halos (upper right panel).
            This may be further evidence that elongated outer stellar haloes are 
            built from accreted stars.
                     
        \item We find no evidence that the rest-frame colour gradients (at $r<60$ kpc) 
            of massive galaxies depend on halo mass. 
            
    \end{itemize}
    
    The fact that we find smooth ellipticity profiles and shallow gradients favors 
    the idea of using a flux-weighted average isophotal shape to extract 1-D \mden{} 
    profiles for massive galaxies. 
    The similarity in the average rest-frame colour profiles demonstrates that our 
    results can not simply be explained by differences in radial \m2l{} ratios between 
    the two samples.
    
    Our work does not address color gradients below 5--6 kpc because we do not 
    deconvolve for the PSF. 
    Color gradients on these scales may be sensitive to other physical processes and 
    deserve future investigation using 2-D modelling methods and/or images with higher 
    spatial resolution.
    
   

\section{Discussion}
    \label{sec:discussion}

\subsection{The Role of Environment in the Two-Phase Formation Scenario}
            
    Using deep images from the HSC survey, we show that the stellar mass distributions 
    of massive central galaxies at $0.3 < z < 0.5$ depend on halo mass. 
    At fixed total galaxy mass, central galaxies in massive halos have larger 
    half-light radii and host more prominent outer stellar haloes compared to galaxies 
    in low mass haloes (Figure~\ref{fig:scaling_relation}). 
    We also find that the outer stellar haloes ($R>50$ kpc) of massive galaxies show 
    the strongest variations with halo mass. 
    
    The two-phase formation scenario can qualitatively explain these results. 
    In this scenario, intense dissipative processes at $z > 2$ are responsible for the 
    formation of \textit{in situ} stars in massive central galaxies. 
    After a rapid quenching of star formation, the subsequent assembly of massive 
    galaxies is dominated by the accretion of satellite galaxies through (mostly) 
    non--dissipative mergers. 
    Dry minor mergers are efficient at depositing \textit{ex situ} stars in the 
    outskirts of massive galaxies (e.g., \citealt{Oogi2013, Bedorf2013}) and hence in 
    building up outer stellar haloes. 
    The fact that minor mergers become more frequent in more massive dark matter 
    haloes could lead to the halo mass dependence of galaxy profiles that we identify 
    in Figure~\ref{fig:prof_1}.
      
    \citet{Shankar2014} studied the environment dependence of galaxy size using 
    semi-analytic models. 
    They predict that at fixed stellar mass, the median size of central galaxies 
    should increase strongly with halo mass. 
    Although the major-merger rate does not strongly depend on halo mass 
    (mass ratio $>1$:3; e.g., \citealt{Hirschmann2013, Shankar2015}), the 
    minor--merger rate could still increase with halo mass if the dynamical friction
    timescale is short (e.g., \citealt{Newman2012}). 
    Massive central galaxies with $11.5 <$\logms{}$<12.0$ living in haloes with 
    \logmh{}$>14.0$ can have up to four times more minor mergers (1:100--1:3) compared 
    to those in less massive haloes at fixed galaxy mass. 
    An increase in the minor-merger rate as a function of halo mass can lead to a 
    halo-mass dependence in the mass--size relation. 
    The predictions from \citet{Shankar2014} have been confirmed by \citet{Yoon2017} 
    using the semi-analytic model from \citet{Guo2011}.  

    Our results are broadly consistent with these predictions. 
    As an important next step, we are using the HSC galaxy-galaxy weak lensing results 
    (\citealt{Mandelbaum2018}) to help us achieve a more detailed picture of the 
    environment dependence. 
    The preliminary result so far has confirmed the trends found in this work with 
    more straightforward and accurate constraints of halo mass (Huang \etal in prep.).
    However, our results alone cannot rule out other explanations for this halo mass 
    dependence. 
    For instance, \citet{Buchan2016} suggests that, under the extreme situation that 
    the majority of the baryons in the halo of the main progenitor can be converted into 
    stars, the in-situ component alone can account for the environment difference 
    we see today. 
    Although very unlikely, it requires comparisons with high redshift observations 
    to distinguish between these two scenarios.

\subsection{The Inner Regions of Massive Galaxies}

    Figure \ref{fig:scaling_relation} shows that, at fixed total galaxy mass, centrals 
    in high mass haloes have slightly shallower inner \mden{} slopes and lower values 
    of \minn{} compared to those in low mass haloes. 
    We have tested that this cannot be solely explained by the choice of a finite 
    aperture (100 kpc) to estimate `total' galaxy mass. 
    Integrating our profiles out to larger radii makes the differences between \rbcg{} 
    and \nbcg{} galaxies on the \mtot{}--\minn{} plane even more significant 
    (see Appendix~\ref{app:robust}). 
    In hydrodynamic simulations, intense dissipative processes help create a 
    self-similar de~Vaucouleurs--like ($n{\sim} 4$) inner density profile 
    (e.g., \citealt{Hopkins2008}). 
    However, there are a variety of physical processes that can shape and alter the 
    inner  profile which we now discuss.
    
    First, major mergers can redistribute the inner stellar mass distributions, but 
    the major--merger rate does not strongly depend on halo mass 
    (e.g., \citealt{Shankar2014}). 
    However, minor mergers, which do depend on halo mass, can also modify central 
    surface brightness profiles (e.g., \citealt{BoylanKolchin2007}). 
    Depending on the structure, and the orbits of infalling satellites, a minor merger 
    can make the inner \mden{} profiles either steeper or shallower.  
    Interestingly, \citet{BoylanKolchin2007} find that when a satellite is accreted 
    from a highly eccentric and energetic orbit to a core--elliptical galaxy, the 
    process tends to reduce the central \mden{}.  
    This is relevant here, as many massive ETGs are known to be core--elliptical 
    galaxies. 
    
    Second, strong adiabatic expansion induced by powerful AGN feedback is 
    another mechanism (e.g., \citealt{Fan2008, Martizzi2013}) that can modify stellar 
    density profiles. 
    When the induced mass loss is efficient enough, it can lead to expanded 
    central stellar mass distribution and can significantly lower the inner \mden{}. 
    
    Finally, the coalescence of super-massive black holes (SMBHs) can also flatten 
    the central \mden{} profile via an efficient scattering effect 
    (e.g., \citealt{Milosavljevi2002}).
    On the right side of Fig \ref{fig:scaling_relation}, there are a few candidates 
    for galaxies with large cores. 
    These may be similar to the recently discovered massive brightest cluster galaxies 
    (BCGs) with very large depleted cores (a few thousand parsecs; e.g., 
    \citealt{Postman2012, LopezCruz2014, Thomas2016, Bonfini2016}) possibly resulting 
    from SMBH mergers.
    
    The impact of these processes on \mden{} profiles and their dependency on 
    halo mass are important questions that warrant further investigation.


\subsection{Comparison with Previous Work} 

    Many previous studies that focused on the mass--size relation found this relation 
    to be independent of halo mass or environment at $z\sim 0.0$ 
    (e.g., \citealt{Nair2010, Maltby2010, Cappellari2013, HCompany13}). 
    Shallow imaging and the use of models which do not necessarily well describe 
    massive galaxies (e.g., single-\ser{} or de~Vaucouleurs models) may have masked 
    the effect revealed in this paper. 
    In Appendix~\ref{app:gama}, we use the \rbcg{} and \nbcg{} galaxies that overlap 
    with the Galaxy And Mass Assembly (GAMA) survey to demonstrate this point. 
    We show that, for galaxies with similar \mstar{} derived from single-\ser{} 
    models using shallower SDSS images (\citealt{Kelvin2012}), the ones in more 
    massive dark matter haloes actually show more prominent outer stellar haloes.
    These types of issues complicate the comparisons of mass--size relations 
    derived from different images or using different methods. 
    For this reason, we only present a qualitative comparison with previous work.
    
    At low redshift, \citet{Cerbrian2014} find ETGs with \logms{}$>11.5$ to be slightly 
    larger in more massive haloes, and they also show that this trend is reversed at 
    lower \mstar{}. 
    \citet{Kuchner2017} use one massive cluster at $z=0.44$ to show that ETGs in 
    that cluster have larger sizes than ETGs in the `field'. 
    \citet{Yoon2017} present a study of a large sample of $z\sim0.1$ SDSS ETGs using 
    a nonparametric method. 
    They also find an environmental dependence of the mass--size relation at 
    \logms{}$>11.2$. 
    Similar to our work, they find that massive ETGs in dense environments are 
    20--40\% larger compared to those in underdense environments.
    Recently, \citet{Charlton2017} use a single-\ser{} model for galaxies at 
    $0.2 < z < 0.8$ in the Canada France Hawaii Lensing Survey (CFHTLenS) 
    (\citealt{Heymans2012}) together with galaxy--galaxy lensing measurements to 
    show that larger galaxies tend to live in more massive dark matter haloes 
    (also see \citealt{Sonnenfeld2017}).
    These results are in broad agreement with those presented here.
    
    As the halo mass dependence of the sizes and \mden{} profiles of massive 
    galaxies is being confirmed at low--redshift, the physical origin and 
    redshift evolution of such dependence will become of increasing interest.  
    Right now, some observations find a strong environmental dependence of the 
    mass--size relation for massive quiescent or early-type galaxies at 
    high--redshift (e.g., \citealt{Papovich2012, Bassett2013, Lani2013, 
    Strazzullo2013, Delaye2014}), while other works suggest otherwise (e.g., 
    \citealt{Rettura2010, Raichoor2012, Kelkar2015, Allen2015}). 
    Comparison with high--redshift results are complicated by many issues and 
    is beyond the scope of this work, but we want to point out again that 
    previous works mostly focus on the mass--size relation while direct comparison 
    of \mden{} profile (e.g., \citealt{Szomoru2012, Patel2013, Buitrago2017, 
    Hill2017}) could help us trace the redshift evolution of the environment 
    dependence better. 
    
\subsection{Towards Consistent Size Definitions}
        
    Until recently, semi-analytic models and hydrodynamic simulations have had
    difficulty reproducing the mass--size relation of massive galaxies. 
    Galaxy sizes are sensitive to many different physical processes (star-formation, 
    feedback, mergers), and matching the galaxy stellar-mass function does not 
    automatically guarantee a match to the mass--size relation. 
    Furthermore, while some effort have been made to use consistent size definitions 
    (\citealt{McCarthy2017}), more often than not, comparisons of the mass--size relation 
    do not use consistent size definitions, or they only perform crude size conversions 
    (e.g., 3-D radii in simulation versus 2-D projected radii in observation; 
    \citealt{Genel2017}). 
    Observers often quote `sizes' corresponding to the half-light radius along the 
    major axis using 2-D projected images. 
    Simulations, on the other hand, often employ sizes that correspond to the 3-D 
    aperture half-mass radius (e.g., \citealt{Price2017}). 
    
    As emphasized earlier, definitions of galaxy `size' in observations are also not 
    always consistent. 
    Measurements of `size' depend on image quality (e.g., seeing, imaging depth), filter, 
    and the adopted method. 
    Although the elliptical single-\ser{} model is widely adopted in measuring the 
    size of galaxies of different types and at different redshifts, it sometimes leads
    to biased results as it does not universally describe all types of galaxies. 
    Sizes derived from 1-D curves-of-growth are more model independent and have been 
    shown to be useful in revealing the environmental dependence of the mass--size 
    relation (e.g., \citealt{Yoon2017}, and this work), but this method does not take 
    the PSF into account. 
    It also depends on imaging depth and background subtraction.

    In this paper, we quote stellar masses measured within elliptical apertures of 
    fixed physical size. 
    We argue that this approach will allow for a more straightforward comparison 
    between observations and theoretical predictions. 
    Even better, we argue that galaxy mass profiles can be compared directly with 
    predictions from hydrodynamic simulations, bypassing the need completely for 
    `size' estimates.
          


\section{Summary and Conclusions}
    \label{sec:summary}
    
    In this paper, we investigate how the stellar mass profiles of massive galaxies 
    depend on the masses of their host dark matter haloes. 
    Using high-quality images from the first $\sim100$ deg$^2$ of the Hyper Suprime-Cam 
    Subaru Strategic Program, we divide central galaxies at $0.3 < z < 0.5$ into two 
    samples according to dark matter halo mass (\mhalo{}$\simgt 10^{14.2} M_{\odot}$ 
    and \mhalo{}$\simlt 10^{14} M_{\odot}$). 
    Exquisite data from HSC enables us to extract stellar profiles for individual 
    galaxies in these two samples out to 100 kpc.
    
    Our main results are as follows:  
    
    \begin{enumerate}
    
    \item At fixed \mtot{}, central galaxies in high mass haloes display shallower 
        \mden{} profiles compared to those in low mass haloes: they have flatter inner 
        \mden{} profiles and more significant outer stellar envelopes.  
        This trend is most pronounced at $R>50$ kpc and thus would easily be missed 
        with shallow imaging data.                 
                                
    \item Massive galaxies matched by \minn{} display a range of \mtot{} values and a 
        large intrinsic scatter in the amplitude of their outer stellar envelopes. 
        
    \item This environmental dependence is also reflected in the mass-size relation, 
        as well as in the \mtot{}-\minn{} relation. 
        We propose that simple elliptical aperture masses such as \mtot{} and\minn{} 
        are better statistics to summarize the properties of galaxy profiles than 
        commonly used `size' estimates such as $R_e$.
      
    \item At fixed \minn{}, galaxies in massive halos have more elliptical outer 
        stellar halos compared to galaxies in low mass halos. 
        This may be further evidence that elongated outer stellar haloes are built 
        from accreted stars.
                     
    \item At fixed galaxy mass and at $r<60$ kpc, the rest-frame colour gradients 
        of massive galaxies do not depend on dark matter halo mass. 
             
    \end{enumerate}
    
    These results highlight the importance of deep, high-quality images for studying 
    the assembly of massive dark matter haloes and their central galaxies. 
    Future work will focus on a comparison between our data and predictions from 
    various hydrodynamic simulations. 
    This will enable us to gain further insight into the physical mechanisms that 
    drive the trends discovered in this paper.

\section*{Acknowledgements}

  The authors thank Frank van~den~Bosch for insightful discussions and Shun Saito for 
  helping us estimate the fraction of satellite galaxies in our sample.
  We also thank Felipe Ardilla and Christopher Bradshaw for useful comments. 
  SH thanks Feng-Shan Liu for sharing the \mden{} profile of the $z\sim1$ BCG from 
  his work. 
  This material is based upon work supported by the National Science Foundation under 
  Grant No. 1714610.

  The Hyper Suprime-Cam (HSC) collaboration includes the astronomical communities of 
  Japan and Taiwan, and Princeton University.  The HSC instrumentation and software were
  developed by National Astronomical Observatory of Japan (NAOJ), Kavli Institute
  for the Physics and Mathematics of the Universe (Kavli IPMU), University of Tokyo,
  High Energy Accelerator Research Organization (KEK), Academia Sinica Institute
  for Astronomy and Astrophysics in Taiwan (ASIAA), and Princeton University.  
  Funding was contributed by the FIRST program from Japanese Cabinet Office,  Ministry 
  of Education, Culture, Sports, Science and Technology (MEXT), Japan Society for 
  the Promotion of Science (JSPS), Japan Science and Technology Agency (JST), Toray 
  Science Foundation, NAOJ, Kavli IPMU, KEK, ASIAA, and Princeton University.
   
  Funding for SDSS-III has been provided by Alfred P. Sloan Foundation, the 
  Participating Institutions, National Science Foundation, and U.S. Department of
  Energy. The SDSS-III website is http://www.sdss3.org.  SDSS-III is managed by the
  Astrophysical Research Consortium for the Participating Institutions of the SDSS-III
  Collaboration, including University of Arizona, the Brazilian Participation Group,
  Brookhaven National Laboratory, University of Cambridge, University of Florida, the
  French Participation Group, the German Participation Group, Instituto de Astrofisica
  de Canarias, the Michigan State/Notre Dame/JINA Participation Group, Johns Hopkins
  University, Lawrence Berkeley National Laboratory, Max Planck Institute for
  Astrophysics, New Mexico State University, New York University, Ohio State University,
  Pennsylvania State University, University of Portsmouth, Princeton University, the
  Spanish Participation Group, University of Tokyo, University of Utah, Vanderbilt
  University, University of Virginia, University of Washington, and Yale University.
  
  The Pan-STARRS1 surveys (PS1) have been made possible through contributions of  
  Institute for Astronomy; University of Hawaii; the Pan-STARRS Project Office; 
  the Max-Planck Society and its participating institutes: the Max Planck Institute 
  for Astronomy, Heidelberg, and the Max Planck Institute for Extraterrestrial Physics, 
  Garching; Johns Hopkins University; Durham University; University of Edinburgh; 
  Queen's University Belfast; Harvard-Smithsonian Center for Astrophysics; Las 
  Cumbres Observatory Global Telescope Network Incorporated; National Central 
  University of Taiwan; Space Telescope Science Institute; National Aeronautics 
  and Space Administration under Grant No. NNX08AR22G issued through the Planetary 
  Science Division of the NASA Science Mission Directorate; National Science 
  Foundation under Grant No. AST-1238877; University of Maryland, and Eotvos 
  Lorand University. 
  
  This paper makes use of software developed for the Large Synoptic Survey 
  Telescope. We thank the LSST project for making their code available as free 
  software at http://dm.lsstcorp.org.
 
  This research was supported in part by National Science Foundation under Grant 
  No. NSF PHY11-25915. 
  
  This research made use of:
  \href{http://www.stsci.edu/institute/software_hardware/pyraf/stsci\_python}{\texttt{STSCI\_PYTHON}},
      a general astronomical data analysis infrastructure in Python. 
      \texttt{STSCI\_PYTHON} is a product of the Space Telescope Science Institute, 
      which is operated by Association of Universities for Research 
      in Astronomy (AURA) for NASA;
  \href{http://www.scipy.org/}{\texttt{SciPy}},
      an open source scientific tool for Python (\citealt{SciPy});
  \href{http://www.numpy.org/}{\texttt{NumPy}}, 
      a fundamental package for scientific computing with Python (\citealt{NumPy});
  \href{http://matplotlib.org/}{\texttt{Matplotlib}}, 
      a 2-D plotting library for Python (\citealt{Matplotlib});
  \href{http://www.astropy.org/}{\texttt{Astropy}}, a community-developed 
      core Python package for astronomy (\citealt{AstroPy}); 
  \href{http://scikit-learn.org/stable/index.html}{\texttt{scikit-learn}},
      a machine-learning library in Python (\citealt{scikit-learn}); 
  \href{http://www.astroml.org/}{\texttt{astroML}}, 
      a machine-learning library for astrophysics (\citealt{astroML});
  \href{https://ipython.org}{\texttt{IPython}}, 
      an interactive computing system for Python (\citealt{IPython});
  \href{https://github.com/kbarbary/sep}{\texttt{sep}} 
      Source Extraction and Photometry in Python (\citealt{PythonSEP});
  \href{https://jiffyclub.github.io/palettable/}{\texttt{palettable}},
      colour palettes for Python;
  \href{http://dan.iel.fm/emcee/current/}{\texttt{emcee}}, 
      Seriously Kick-Ass MCMC in Python;
  \href{http://bdiemer.bitbucket.org/}{\texttt{Colossus}}, 
      COsmology, haLO and large-Scale StrUcture toolS (\citealt{Colossus}).

\bibliographystyle{mnras}
\bibliography{redbcg}

\clearpage
\include{table1}
\clearpage


\appendix

\section{Basic Statistical Properties of the Sample} 
	\label{app:basic} 
    
  \begin{figure}
      \centering 
      \includegraphics[width=8.5cm]{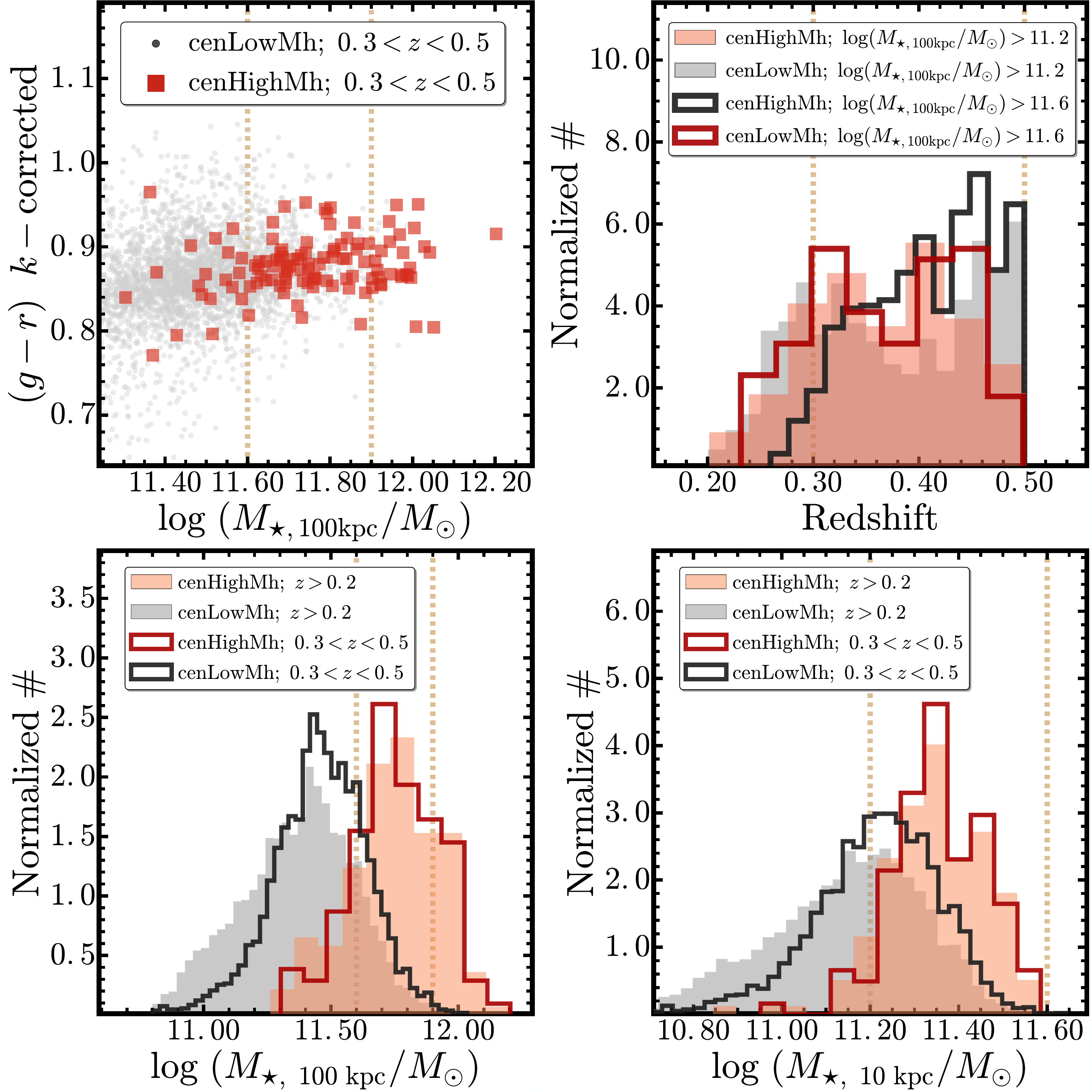}
      \caption{
          \textbf{Top-left}: The \logms{}-$g-r$ colour relation of the \rbcg{} 
          (red circle) and \nbcg{} (grey dots) samples.
          We apply the $k$-corrections from \texttt{iSEDFit} fitting to the colours.~~          
          \textbf{Top-right}: the histograms of the redshift for the \rbcg{} and 
          \nbcg{} galaxies in both \logmtot$>11.2$ and $11.6<$\logmtot{}$<11.9$
          mass bins.
          The vertical lines highlights the $0.3\leq z \leq 0.5$ redshift range.~~
          \textbf{Bottom-left}: the histograms of \mtot{} for the \rbcg{} (orange-red) 
          and \nbcg{} (grey-black) samples at both $z>0.2$ (step-filled histogram) and 
          $0.3 \leq z \leq 0.5$ (stepped histogram). 
          The vertical lines in both top-left and bottom-left figures highlight the 
          $11.6<$\logmtot{}$<11.9$ mass range that will be used in the comparison of 
          the \mtot{}-matched samples.~~
          \textbf{Bottom-right}: the histograms of \minn{} in similar format. 
          Here the vertical lines highlight the 
          $11.2<$\logminn{}$<11.6$ mass range that is used for comparison.
      }
      \label{fig:sample_stats}
  \end{figure}

    On the top-left panel of Fig~\ref{fig:sample_stats}, we show the \mtot{}-colour 
    relations using the $k$-corrected $g-r$ colour. 
    Both samples follow the same tight `red-sequence' with little contamination 
    from the `blue cloud'.
    At fixed \mtot{}, we see little offset in colour distributions of the two 
    samples, suggesting that both samples consist of quiescent galaxies with 
    similar average stellar population properties.  
    This is consistent with previous result that suggests the average stellar 
    population of massive central galaxy does not depend on \mhalo{} 
    (e.g.\ \citealt{Park2007}).  
    In this work, we focus on the \mstar{} range of $11.6 \le$\logmtot{}$\le 11.9$, 
    where both samples have acceptable completeness, and their \mtot{} distributions 
    greatly overlap (see the normalized distributions of \mtot{} in the bottom-left 
    panel of Fig~\ref{fig:sample_stats}). 
    As for the \minn{} distributions, the two samples overlap the most within 
    $11.2 \le$\logmtot{}$\le 11.6$, but now they show quite different
    distributions (bottom-right figure).
    
    The redshift distributions also show small difference
    (upper-right panel) even in the high-\mtot{} bin, where the redshift distribution 
    of the \nbcg{} sample skews toward higher-$z$ end due to the contribution of BOSS 
    spec-$z$.
    Since this could bias the comparison of \mden{} profiles and other properties 
    (please see Appendix\ref{app:redshift} for more details), we address this via matching 
    the two samples in both mass and redshift distributions carefully
    (see Appendix~\ref{app:match}).

\section{Comparisons of \mden{} profiles in different redshift bins}
    \label{app:redshift}
    
\begin{figure}
    \centering 
    \includegraphics[width=8.2cm]{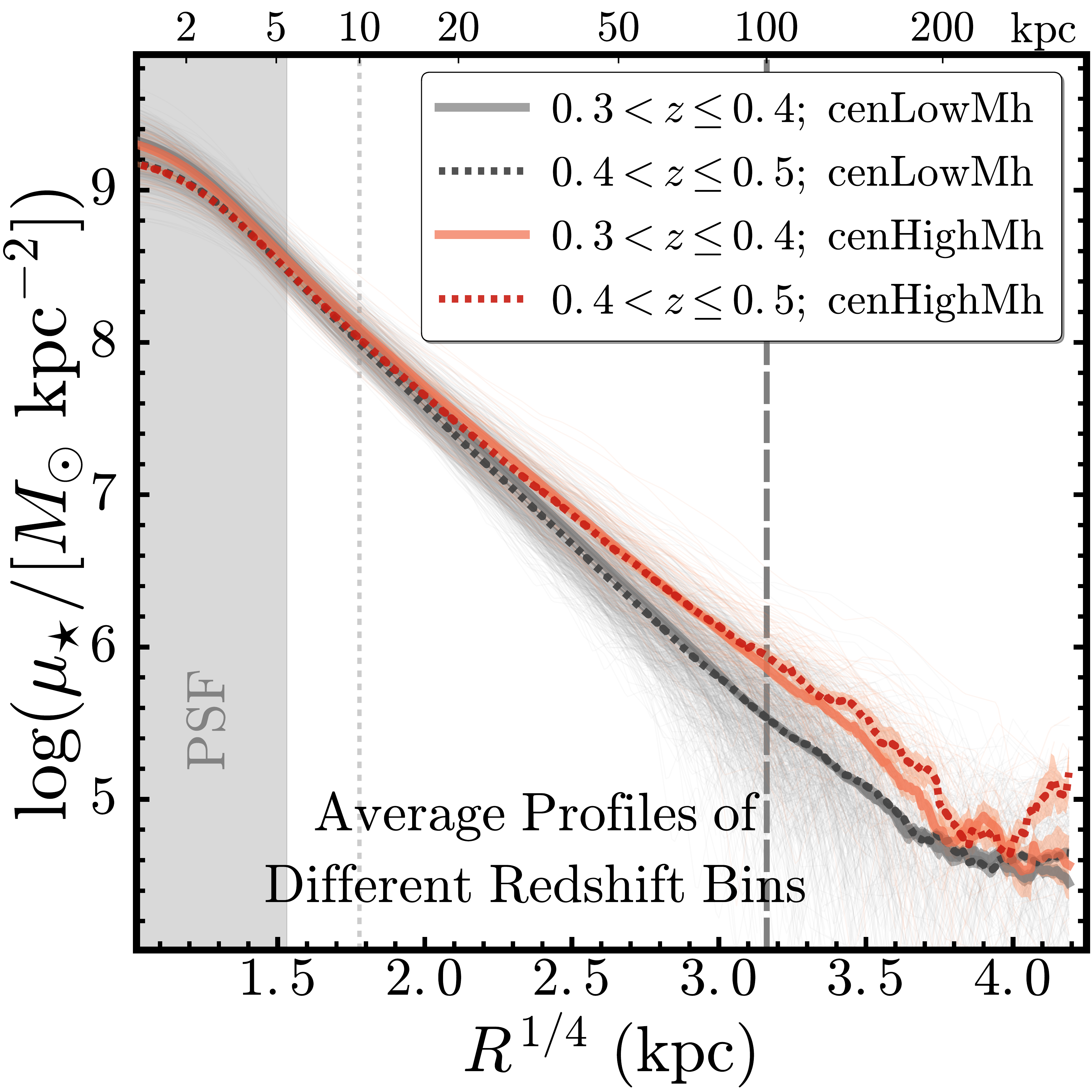}
    \caption{
        Comparison of \mden{} profiles of \rbcg{} (orange-red) and \nbcg{} 
        (grey-black) at $11.6 \le$\logmtot$< 11.9$ in redshift bins of 
        $0.3\leq z<0.4$ (solid lines) and $0.4\leq z<0.5$ (dash lines). 
        We show the individual profile in the background using much thinner line, 
        and highlight the median profiles using thicker line and darker colour.
        }
    \label{fig:avg_prof_z}
\end{figure}    

\begin{figure*}
    \centering 
    \includegraphics[width=15.0cm]{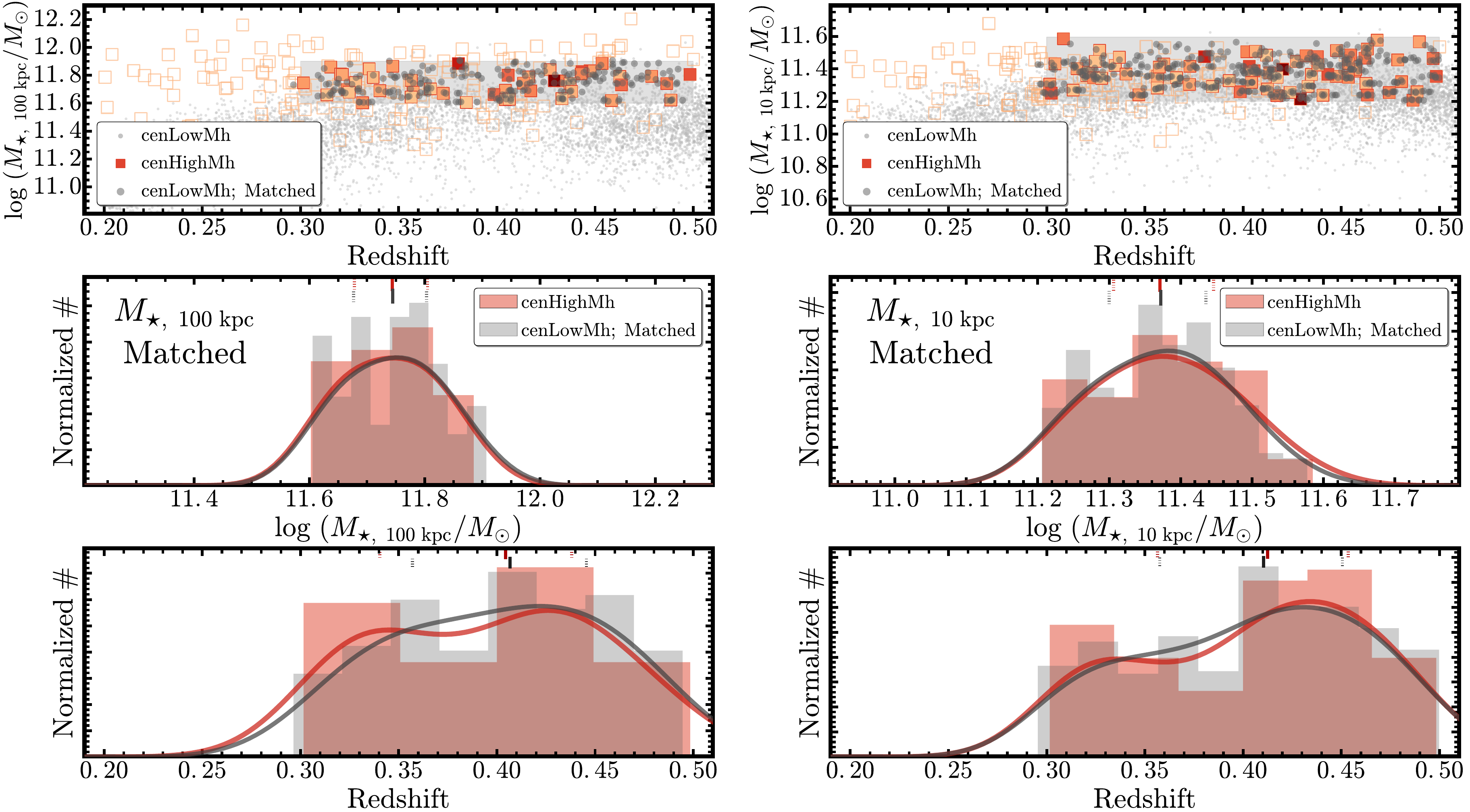}
    \caption{
        \textbf{Left figure} shows the details of the \mtot{}-matching process, 
        corresponding to the results shown in the left figure of   
        On the top panel, we show the overall distributions of \rbcg{} (light orange boxes) 
        and \nbcg{} (light grey dots) galaxies on the \mtot{}-$z$ plane.  
        And, we match the two sample in the \mtot{}-$z$ space outlined by the shaded region.
        We highlight the \rbcg{} galaxies in this region using bigger boxes in red frames, 
        whose size reflects the $P_{\mathrm{Cen}}$ value.  
        We also colour-code them using the richness ($\lambda$) of the host cluster. 
        The matched \nbcg{} galaxies are highlighted using darker colour and bigger dots. 
        To further evaluate the matching results, we show the distributions of \mtot{} 
        (middle panel) and redshift (bottom panel) separately. 
        On both panels, we show the histograms along with their kernel density 
        distributions.  
        And, on the top of each panel, two sets of short vertical lines highlight the median 
        value (solid) and the inter-quartile (dash) of each distribution.~~~
        \textbf{Right figure} shows the similar matching results for the \minn{}-matched
        samples used for the right figure of Fig~\ref{fig:prof_1}.
        The format is exactly the same as the left one, except the \minn{} replaces the 
        \mtot{} in the top and middle panels.}
    \label{fig:match}
\end{figure*}

    Given the redshift range for our samples, it is important to evaluate 
    the impacts from the physical extend of seeing and the imaging depth on the \mden{} 
    profiles at different redshift. 
    Under the same seeing, the \mden{} profile of galaxy at higher-$z$ is more 
    vulnerable to the PSF smearing effect at the center. 
    It is also harder to reach to the same \mden{} level under the same imaging depth 
    due to cosmological dimming and background noise. 
    
    In Fig~\ref{fig:avg_prof_z}, we group the \rbcg{} and \nbcg{} galaxies within 
    $11.6 \le$\logmtot$< 11.9$ into two $z$ bins ($0.3\leq z<0.4$ and $0.4\leq z<0.5$),
    and compare their \mden{} profiles. 
    In two redshift bins, the median \mden{} profiles from the same sample follow each 
    other very well outside 10 kpc, but become visibly different in the central 3-4 kpc,
    where the effect from seeing kicks in. 
    Meanwhile, the median \mden{} profiles of \rbcg{} and \nbcg{} in the same $z$ bin 
    are identical in the central region, which indicates similar average seeing 
    conditions.       
    This confirms that \mden{} profile at $> 5$ kpc is safe from the impacts of seeing 
    and difference in redshift.
    More importantly, it also suggests that, once the redshift distributions are 
    carefully matched, the difference of \mden{} profile is likely to be physical 
    even in the central region.  

\section{Match the \rbcg{} and \nbcg{} samples in \mstar{} and redshift distributions}
    \label{app:match} 


    As explained earlier, it is important to make sure the two samples have similar 
    distributions in both \mstar{} and redshift before comparing their median \mden{} 
    profiles.  
    Here we briefly describe the procedure used in this work. 
    Since the \rbcg{} sample is smaller in size, we always match the \nbcg{} sample to 
    it by searching for the $N$-nearest neighbours on the $M_{\star}$-redshift plane 
    using the KDTree algorithm in the \texttt{scikit-learn} Python library 
    (\citealt{scikit-learn}), and evaluate the quality of the match using the 
    distributions of both parameters (as shown in Fig~\ref{fig:match}). 
 
    As we only keep the unique \nbcg{} galaxies in the matched sample, we manually 
    adjust the value of $N$ to achieve the best match. 
    When the redshift distribution of the \rbcg{} sample becomes bi-model, we also try 
    to split the sample into two redshift bins and match them separately. 
    Typically $N$ is between 3 to 8.
    In Fig~\ref{fig:match}, we demonstrate this procedure using the results for 
    the \mtot{}-matched (Left) and the \minn{}-matched samples in Fig~\ref{fig:prof_1}
    (Right), and the two samples are well matched in the distributions of \mtot{}
    (or \minn{}) and redshift.  
    For all the comparisons of \mden{} profiles in this work, we match the samples 
    in the same way, and make sure the match has the same quality. 

\section{Robustness of the \mden{} Differences} 
	\label{app:robust}

  \begin{figure*}
      \centering 
      \includegraphics[width=15.5cm]{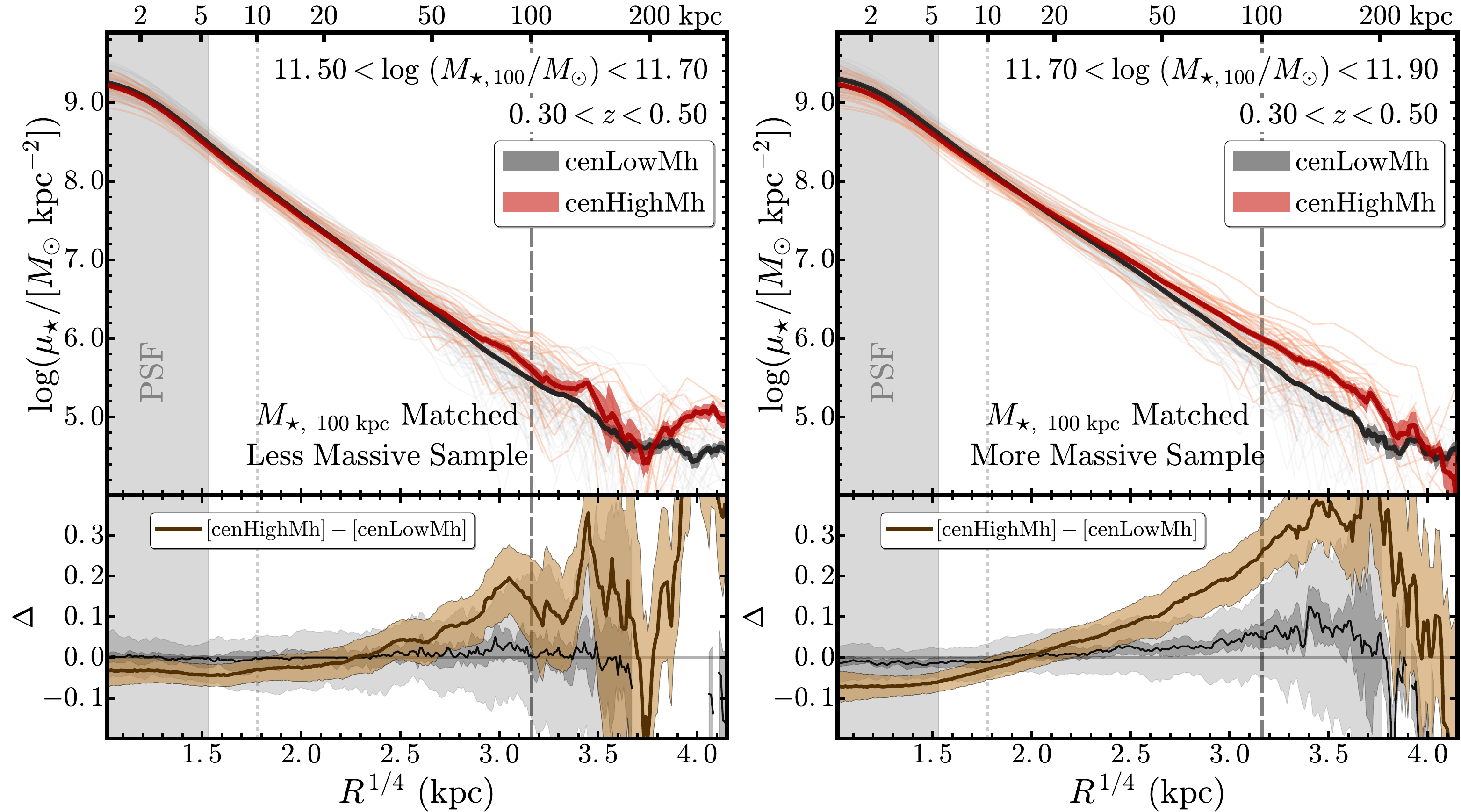}
      \caption{
          Comparisons of the \mden{} profiles for \mtot{}-matched \rbcg{} 
          (orange-red) and \nbcg{} (grey-black) galaxies in lower (left; [11.5,11.70]) 
          and higher (right; [11.7, 11.9]) \mtot{} bins. 
          Other formats are in consistent with the right figure of Fig~\ref{fig:prof_1}.
          The difference in median profiles is more significant in higher \mtot{} bin.
          }
      \label{fig:prof_2}
  \end{figure*}

  \begin{figure*}
      \centering 
      \includegraphics[width=\textwidth]{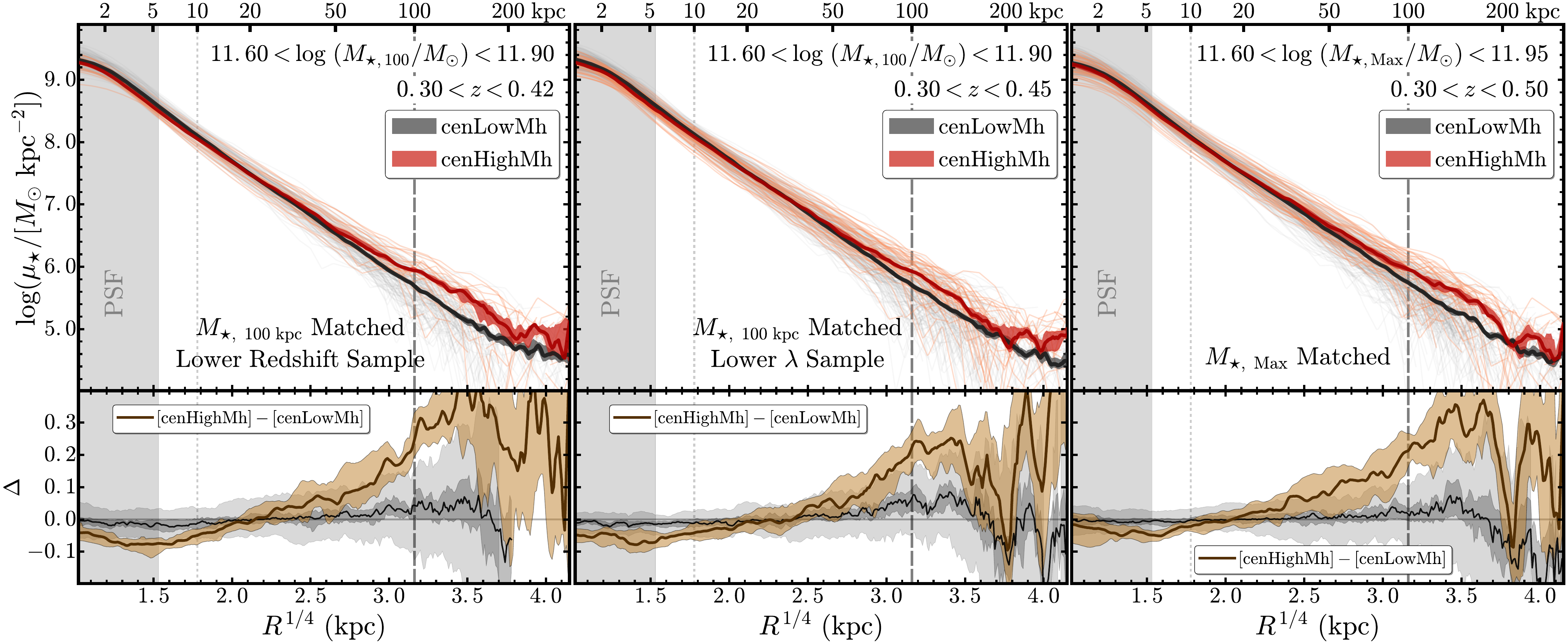}
      \caption{
        Comparisons of the \mden{} profiles for \rbcg{} (orange-red) and \nbcg{} 
      	(grey-black) galaxies that are matched using proxies of total \mstar{}. 
        The formats are in consistent with the right figure of Fig~\ref{fig:prof_1}.
        The differences are, here, the samples are matched in slightly different ways. 
        From left to right: a) using samples at lower redshift ($0.3 < z < 0.4$); 
        b) using \rbcg{} sample with $\lambda > 20$ instead of 30; 
        c) using \mstar{} within 150 kpc instead of 100 kpc.
        The results are broadly consistent with the one in Fig~\ref{fig:prof_1}.
        }
      \label{fig:prof_3} 
  \end{figure*}

  \begin{figure*}
      \centering 
      \includegraphics[width=15.5cm]{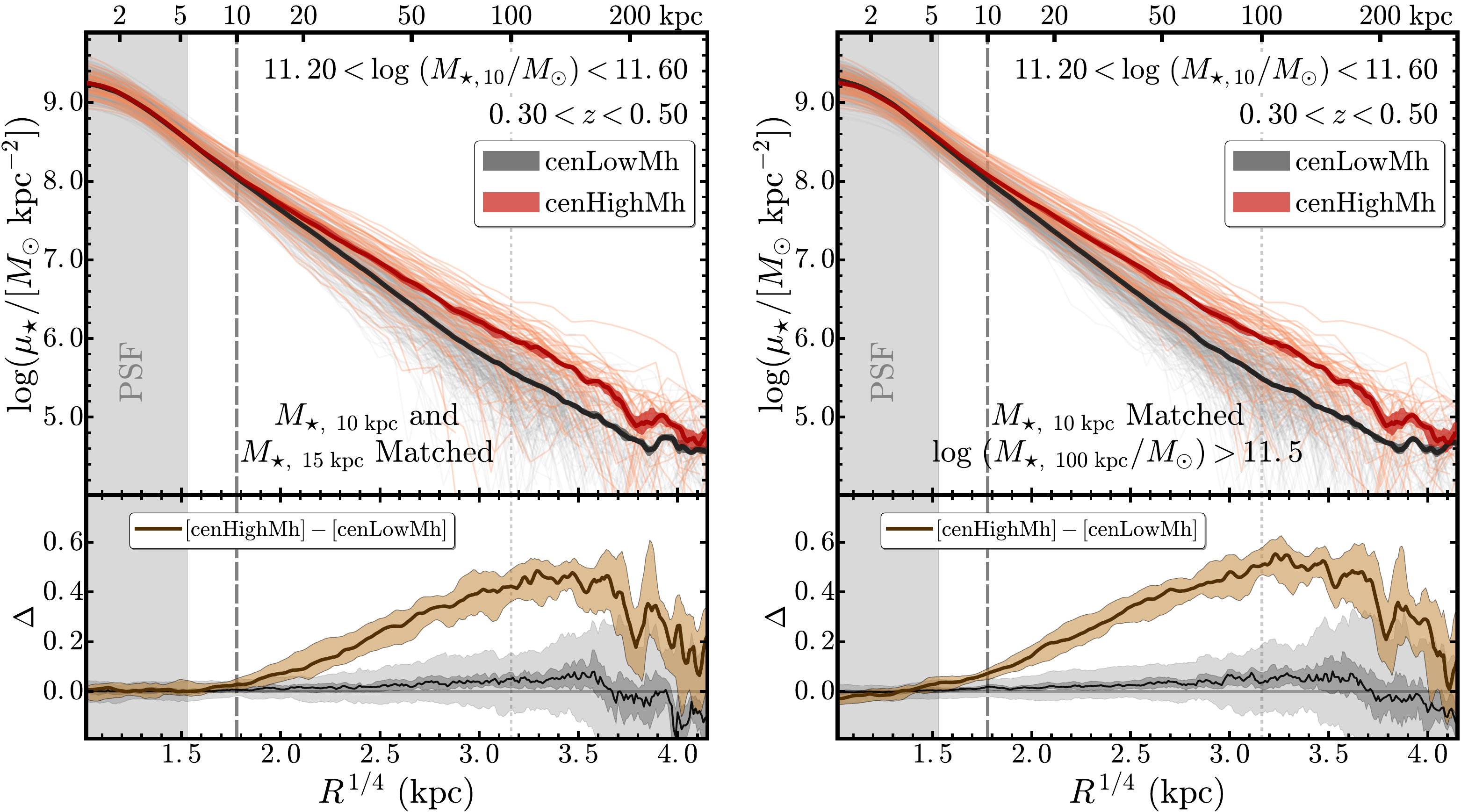}
      \caption{
          Comparisons of the \mden{} profiles for \rbcg{} (orange-red) and \nbcg{} 
          (grey-black) galaxies that are matched using the \mstar{} enclosed in the 
          inner region. 
          Left panel shows the results after matching the \minn{} and \meff{} together, 
          and the right panel shows the results when only the \logmtot{}$\ge 11.5$
          \rbcg{} and \nbcg{} galaxies are included.
          Other formats are in consistent with the right figure of Fig~\ref{fig:prof_1}.
          }
      \label{fig:prof_4} 
  \end{figure*}

    In Fig~\ref{fig:prof_1}, we compare the \mden{} profiles of \mtot{}- and 
    \minn{}-matched samples of \rbcg{} and \nbcg{} galaxies, and here we test the 
    robustness of the results using a few extra tests that are illustrated in
    Fig~\ref{fig:prof_2}, Fig~\ref{fig:prof_3}, and Fig~\ref{fig:prof_4}, and 
    are briefly described here:   
    
    \begin{enumerate}
        
        \item In Fig~\ref{fig:prof_2}, we group the samples into two \mtot{} bins. 
            Given the small sample size, we extend slightly toward lower \mtot{} range 
            ($11.5 \leq \log_{10} (M_{\star,\ 10\mathrm{kpc}}/M_{\odot}) < 11.7$ and 
             $11.7 \leq \log_{10} (M_{\star,\ 10\mathrm{kpc}}/M_{\odot}) < 11.9$). 
            Although the smaller sample leads to larger statistical uncertainties, 
            we can still see similar structural differences in both \mtot{} bins, 
            and the difference becomes more significant in the higher \mtot{} bin.  
            For the lower \mtot{} bin, the difference in the inner region becomes 
            quite uncertain, while the difference in the outskirt is still visible. 
            This potentially suggests that the environmental dependence of structure 
            also varies with \mstar{}, an important implication deserves more 
            investigations in the future.   

        \item On the left panel of Fig~\ref{fig:prof_3}, we match the \rbcg{} and 
            \nbcg{} samples in a lower redshift bins ($0.30 < z < 0.42$).
            Despite the larger uncertainties due to smaller samples, we find the 
            results are the same.
            
        \item On the middle panel of Fig~\ref{fig:prof_3}, we includes \rbcg{} 
            galaxies in poorer clusters ($20 < \lambda < 30$), which should result 
            in overlapped \mhalo{} distributions with the \nbcg{} samples 
            considering the typical uncertainty of $\lambda$.
            This makes the difference in the inner region slightly less significant, 
            but the overall results are the same. 
             
        \item On the right panel of Fig~\ref{fig:prof_3}, in stead of using \mtot{}, 
            we use the \mmax{}--the maximum \mstar{} by integrating the \mden{} 
            profiles to the largest radius allowed.  
            The \mmax{} values are less reliable than \mtot{} due to the 
            uncertainty of background subtraction and contamination from nearby 
            bright objects, but they can serve as different estimates of the `total'
            \mstar{} of these galaxies.
            As shown in \S~\ref{sec:discussion}, they on average increase
            the \mstar{} by a little bit and affect the \rbcg{} more.
            The differences in the \mden{} profiles still remain very similar.
      
    \end{enumerate}
    
    We also test the robustness of the \mtot{}-matched results using the samples with 
    only spectroscopic redshift, the samples in the three GAMA fields, and the \rbcg{} 
    samples without the ones in very massive haloes ($\lambda > 40$).  
    Limited by space, we do not show these results here, but they all verify the 
    robustness of the results. 
    
    For the results from the \minn{}-matched samples: 
    
    \begin{enumerate}
    
        \item
            We match the two samples using both \minn{} and the \mstar{} within 15 kpc 
            at the same time.  
            This makes the two median \mden{} profiles very similar inside 10-15 
            kpc, while the result in the outskirt remains the same (left panel of 
            Fig~\ref{fig:prof_4}).
            Use \mstar{} within 5 or 20 kpc leads to the same conclusion. 
          
        \item 
            To make sure the two samples are comparable in their overall assembly history,
            we also try to only include the very massive galaxies (\logmtot{}$>11.5$)
            in both samples. 
            This excludes the \nbcg{} galaxies that are much less massive and 
            more `compact' in structure. 
            Yet, the results regarding the structural differences remain the same. 
          
    \end{enumerate}

\section{Stellar Mass Curve-of-Growth for Massive Central Galaxies} 
	\label{app:cog}
	
	In \S \ref{ssec:sbp_mtot} and Figure \ref{fig:prof_1}, we show the comparisons 
	of \mden{} profiles for massive central galaxies from the \rbcg{} and \nbcg{} 
	samples. 
	Although the differences in their median \mden{} profiles we revealed are robust 
	and systematic, they appear to be very subtle, especially in the inner region. 
	This is partly due to the logarithmic scale on the Y--axis for \mden{} profiles. 
	
	In Figure \ref{fig:cog}, we compare the same two samples after converting the 
	\mden{} profiles into: 
	
	\begin{enumerate}
	
	    \item `Curve-of-growth' of stellar mass -- the cumulative \logms{} profiles
	        (upper panel).
	    
	    \item Fraction of \mtot{} within different radius (lower panel).
	
	\end{enumerate}
	
	These comparisons demonstrate the same results from different angles and 
	the systematic differences become more clear using the fraction of \mtot{}
	within different radius.  
	The comparison of cumulative \logms{} profiles also demonstrates that the 
	\rbcg{} and \nbcg{} samples have very similar median \mtot{}.  
    They help confirm that the distributions of stellar mass within 100 kpc indeed 
    have systematic differences between the massive central galaxies living in more 
    and less massive dark matter haloes. 


\begin{figure*}
    \centering
    \includegraphics[width=16cm]{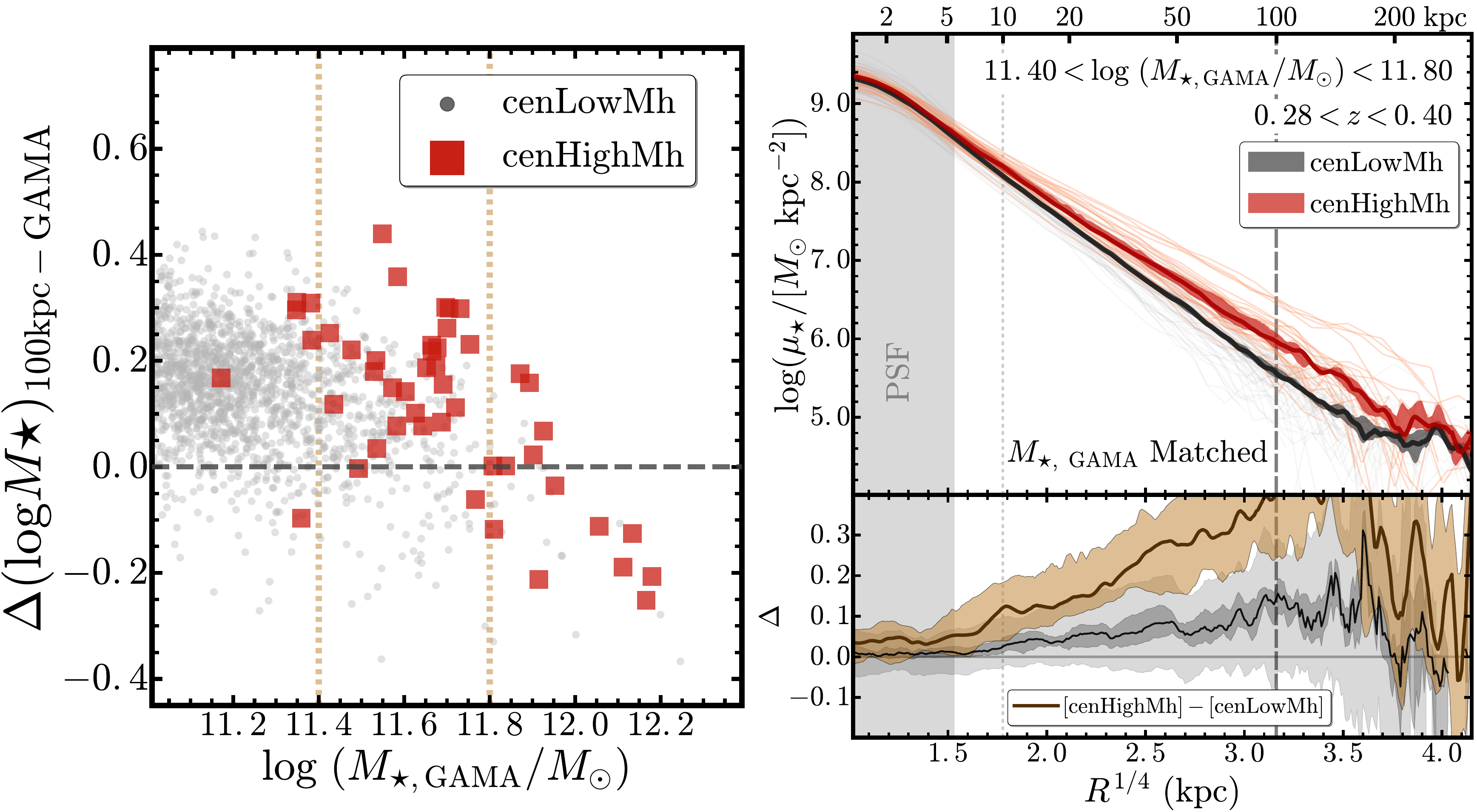}
    \caption{
        \textbf{Left:} comparison of \mstar{} estimated by the GAMA survey and 
        the \mtot{} using HSC images in this work. 
        We plot the \logmgama{} against the difference between \logmtot{} and \logmgama{}. 
        The two vertical lines highlight the mass range $11.4 \leq$\logmgama{}$<11.8$ 
        that is used for the comparison.~~
        \textbf{Right:} we compare the \mden{} profiles of \rbcg{} (orange-red) and 
        \nbcg{} (grey-black) galaxies using the samples matched on the 
        \mgama{}-$z$ plane at $11.4 \leq$\logmgama{}$<11.9$ and $0.28 \leq z < 0.4$. 
        The format is very similar to the ones in Fig~\ref{fig:prof_1}.}
    \label{fig:gama}
\end{figure*}

\section{Comparison of \mden{} profiles using \mstar{} from the GAMA survey}
    \label{app:gama} 

\begin{figure}
    \centering
    \includegraphics[width=\columnwidth]{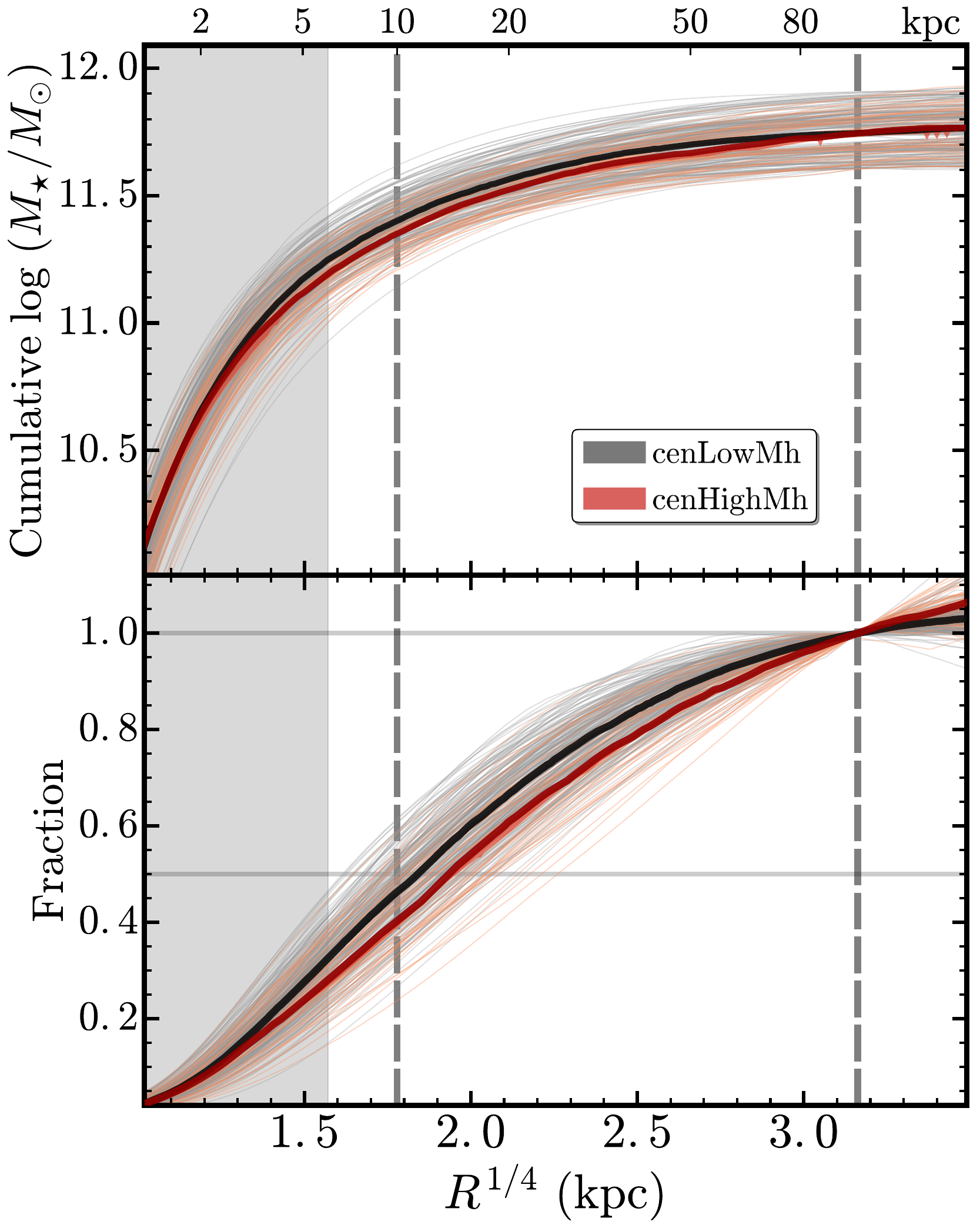}
    \caption{
        \textbf{Top:} comparison of the cumulative \mstar{} profiles for 
        \rbcg{} (orange--red) and \nbcg{} (grey--black) galaxies at fixed \mtot{}.~~
        \textbf{Bottom:} comparison of the fraction of \mtot{} within different radius 
        for the same samples at fixed \mtot{}. 
        Thicker and darker solid lines highlight the median profiles. 
        Other formats of the figure are similar to Figure \ref{fig:prof_1}. 
        Besides the region affected by seeing, the 10 and 100 kpc radius, we 
        also highlight the 50\% and 100\% values using horizontal grey lines on the 
        bottom panel. 
        }
    \label{fig:cog}
\end{figure}
    
    The GAMA survey greatly overlaps with the HSC survey, and it provides carefully 
    measured \mstar{} for large sample of galaxies (\citealt{Taylor2011}) that help 
    produce many interesting results (e.g., \citealt{Bauer2013, Ferreras2017}).
    They use 2-D single-\ser{} model to correct the total luminosity of the galaxy 
    (\citealt{Kelvin2012}), and derive the \m2l{} through optical-SED fitting 
    (BC03 model; Chabrier IMF) based on the PSF-matched aperture photometry. 
    Since the \ser{} model is generally more flexible than the \texttt{cModel} one, 
    it is therefore interesting to compare with the \rbcg{} and \nbcg{} galaxies 
    that also have spec-$z$ (at $z < 0.40$) and \mstar{} in GAMA DR2 
    (\citealt{Liske2015}) and see the impact of deep photometry again. 
    
    We summarize the results in Fig~\ref{fig:gama}.  
    On the left panel, we compare the differences between \mtot{} and \mgama{}. 
    HSC survey on average recovers more \mstar{} at high-\mstar{} end, which is 
    consistent with the expectation from deeper photometry, although the 
    systematic differences in the estimates of \m2l{} could play a role here. 
    Meanwhile, it is interesting see that, above \logmtot{}$> 11.8$, \mgama{} 
    becomes increasingly larger than \mtot{}, and most of these massive 
    galaxies have very high \ser{} index from the 2-D fitting. 
    This suggests that the single-\ser{} model is no longer an appropriate one to 
    describe very massive galaxies as it tends to over-estimate the \mstar{} the 
    inner and/or outer regions. 
    
    To verify the cause of the difference in \mstar{}, we further select samples 
    of \rbcg{} and \nbcg{} galaxies with matched \mgama{} and redshift 
    distributions (at $11.4 <$\logmgama{}$<11.8$; see Appendix \ref{app:match}), 
    and compare their \mden{} profiles (right panel). 
    Although these two subsamples are equally massive according to results from 
    GAMA survey, it is clear that the \rbcg{} galaxy has much more extended 
    outer envelope, even though its median \mden{} profile is very similar 
    to the \nbcg{} sample at $< 10$ kpc. 
    We can reproduce very the same trend with the luminosity density profiles 
    (with or without $k$-correction), suggesting that the inaccurate \ser{} 
    model definitely leads to under-estimate of \mstar{}.  
 

\bsp
\label{lastpage}
\end{document}

%% file: table1.tex
\begin{deluxetable}{c ccc cc cc}
    \label{table:1}
    \centering
    
    \tabletypesize{\scriptsize}
    \tablewidth{0pt}
    \tablecolumns{8}
    \tablenum{1}
    \tablecaption{Average \mden{} Profiles of Massive Galaxies in Different Stellar Mass Bins}
\tablehead{
    \colhead{Radius} & 
    \multicolumn{3}{c}{[\mden{}]; Combined samples} &
    \multicolumn{2}{c}{[\mden{}]; $M_{\star,100\ \mathrm{kpc}}$-matched} &
    \multicolumn{2}{c}{[\mden{}]; $M_{\star,10\ \mathrm{kpc}}$-matched}
	\vspace{1.4ex}
    \nl 
    \colhead{kpc} & 
    \multicolumn{3}{c}{$\log (M_{\odot}/\mathrm{kpc}^2)$} &
    \multicolumn{2}{c}{$\log (M_{\odot}/\mathrm{kpc}^2)$} &
    \multicolumn{2}{c}{$\log (M_{\odot}/\mathrm{kpc}^2)$}
	\vspace{1.4ex}
    \nl 
    \colhead{} & 
    \colhead{$\log \frac{M_{\star,100\mathrm{kpc}}}{M_{\odot}}\in$[11.4, 11.6]} & 
    \colhead{[11.6, 11.8]} & 
    \colhead{[11.8, 12.0]}\hspace{2.0ex} & 
    \colhead{\texttt{cenHighMh}} & 
    \colhead{\texttt{cenLowMh}} & 
    \colhead{\texttt{cenHighMh}}\hspace{2.0ex} & 
    \colhead{\texttt{cenLowMh}}
	\vspace{1.6ex}
    \nl
    \colhead{    (1)} &
    \colhead{    (2)} &
    \colhead{    (3)} &
    \colhead{    (4)} &
    \colhead{    (5)} &
    \colhead{    (6)} &
    \colhead{    (7)} &
    \colhead{    (8)}
}
\startdata

0.0 & $ 9.23\substack{+0.00 \\ -0.00}$ &$ 9.31\substack{+0.00 \\ -0.01}$ &$ 9.32\substack{+0.01 \\ -0.01}$ &$ 9.31\substack{+0.02 \\ -0.02}$ &$ 9.34\substack{+0.01 \\ -0.01}$ &$ 9.31\substack{+0.02 \\ -0.02}$ &$ 9.34\substack{+0.02 \\ -0.02}$ \\
 0.6 & $ 9.20\substack{+0.00 \\ -0.00}$ &$ 9.28\substack{+0.00 \\ -0.01}$ &$ 9.29\substack{+0.01 \\ -0.01}$ &$ 9.27\substack{+0.02 \\ -0.02}$ &$ 9.31\substack{+0.01 \\ -0.01}$ &$ 9.28\substack{+0.02 \\ -0.02}$ &$ 9.31\substack{+0.02 \\ -0.02}$ \\
 1.0 & $ 9.16\substack{+0.00 \\ -0.00}$ &$ 9.24\substack{+0.00 \\ -0.00}$ &$ 9.26\substack{+0.01 \\ -0.01}$ &$ 9.24\substack{+0.02 \\ -0.02}$ &$ 9.27\substack{+0.01 \\ -0.01}$ &$ 9.25\substack{+0.02 \\ -0.02}$ &$ 9.27\substack{+0.02 \\ -0.02}$ \\
 1.4 & $ 9.12\substack{+0.00 \\ -0.00}$ &$ 9.20\substack{+0.00 \\ -0.00}$ &$ 9.23\substack{+0.01 \\ -0.01}$ &$ 9.20\substack{+0.02 \\ -0.02}$ &$ 9.23\substack{+0.01 \\ -0.01}$ &$ 9.21\substack{+0.02 \\ -0.01}$ &$ 9.23\substack{+0.02 \\ -0.01}$ \\
 1.7 & $ 9.06\substack{+0.00 \\ -0.00}$ &$ 9.15\substack{+0.00 \\ -0.00}$ &$ 9.19\substack{+0.01 \\ -0.01}$ &$ 9.15\substack{+0.02 \\ -0.02}$ &$ 9.19\substack{+0.01 \\ -0.01}$ &$ 9.16\substack{+0.01 \\ -0.01}$ &$ 9.18\substack{+0.01 \\ -0.01}$ \\
 2.0 & $ 9.00\substack{+0.00 \\ -0.00}$ &$ 9.10\substack{+0.00 \\ -0.00}$ &$ 9.15\substack{+0.01 \\ -0.01}$ &$ 9.09\substack{+0.01 \\ -0.02}$ &$ 9.13\substack{+0.01 \\ -0.01}$ &$ 9.11\substack{+0.01 \\ -0.01}$ &$ 9.12\substack{+0.01 \\ -0.01}$ \\
 2.4 & $ 8.93\substack{+0.00 \\ -0.00}$ &$ 9.03\substack{+0.00 \\ -0.00}$ &$ 9.09\substack{+0.01 \\ -0.01}$ &$ 9.03\substack{+0.02 \\ -0.02}$ &$ 9.07\substack{+0.01 \\ -0.01}$ &$ 9.05\substack{+0.01 \\ -0.01}$ &$ 9.05\substack{+0.01 \\ -0.01}$ \\
 2.7 & $ 8.87\substack{+0.00 \\ -0.00}$ &$ 8.97\substack{+0.00 \\ -0.00}$ &$ 9.04\substack{+0.01 \\ -0.01}$ &$ 8.97\substack{+0.01 \\ -0.01}$ &$ 9.01\substack{+0.01 \\ -0.01}$ &$ 9.00\substack{+0.01 \\ -0.01}$ &$ 8.99\substack{+0.01 \\ -0.01}$ \\
 3.0 & $ 8.80\substack{+0.00 \\ -0.00}$ &$ 8.90\substack{+0.00 \\ -0.00}$ &$ 8.98\substack{+0.01 \\ -0.01}$ &$ 8.90\substack{+0.01 \\ -0.01}$ &$ 8.95\substack{+0.01 \\ -0.01}$ &$ 8.93\substack{+0.01 \\ -0.01}$ &$ 8.92\substack{+0.01 \\ -0.01}$ \\
 3.4 & $ 8.72\substack{+0.00 \\ -0.00}$ &$ 8.83\substack{+0.00 \\ -0.00}$ &$ 8.92\substack{+0.01 \\ -0.01}$ &$ 8.83\substack{+0.01 \\ -0.01}$ &$ 8.88\substack{+0.01 \\ -0.01}$ &$ 8.86\substack{+0.01 \\ -0.01}$ &$ 8.85\substack{+0.01 \\ -0.01}$ \\
 3.7 & $ 8.66\substack{+0.00 \\ -0.00}$ &$ 8.78\substack{+0.00 \\ -0.00}$ &$ 8.87\substack{+0.01 \\ -0.01}$ &$ 8.78\substack{+0.01 \\ -0.01}$ &$ 8.83\substack{+0.01 \\ -0.01}$ &$ 8.81\substack{+0.01 \\ -0.01}$ &$ 8.79\substack{+0.01 \\ -0.01}$ \\
 4.1 & $ 8.60\substack{+0.00 \\ -0.00}$ &$ 8.72\substack{+0.00 \\ -0.00}$ &$ 8.82\substack{+0.01 \\ -0.01}$ &$ 8.72\substack{+0.01 \\ -0.01}$ &$ 8.77\substack{+0.01 \\ -0.01}$ &$ 8.76\substack{+0.01 \\ -0.01}$ &$ 8.73\substack{+0.01 \\ -0.01}$ \\
 4.4 & $ 8.54\substack{+0.00 \\ -0.00}$ &$ 8.66\substack{+0.00 \\ -0.00}$ &$ 8.77\substack{+0.01 \\ -0.01}$ &$ 8.66\substack{+0.01 \\ -0.01}$ &$ 8.72\substack{+0.01 \\ -0.01}$ &$ 8.70\substack{+0.01 \\ -0.01}$ &$ 8.67\substack{+0.01 \\ -0.01}$ \\
 4.8 & $ 8.48\substack{+0.00 \\ -0.00}$ &$ 8.60\substack{+0.00 \\ -0.00}$ &$ 8.71\substack{+0.01 \\ -0.01}$ &$ 8.60\substack{+0.01 \\ -0.01}$ &$ 8.66\substack{+0.01 \\ -0.01}$ &$ 8.65\substack{+0.01 \\ -0.01}$ &$ 8.61\substack{+0.01 \\ -0.01}$ \\
 6.2 & $ 8.26\substack{+0.00 \\ -0.00}$ &$ 8.40\substack{+0.00 \\ -0.00}$ &$ 8.53\substack{+0.01 \\ -0.01}$ &$ 8.41\substack{+0.01 \\ -0.01}$ &$ 8.46\substack{+0.01 \\ -0.01}$ &$ 8.46\substack{+0.02 \\ -0.02}$ &$ 8.40\substack{+0.02 \\ -0.02}$ \\
 7.6 & $ 8.09\substack{+0.00 \\ -0.00}$ &$ 8.24\substack{+0.00 \\ -0.00}$ &$ 8.39\substack{+0.01 \\ -0.01}$ &$ 8.27\substack{+0.01 \\ -0.01}$ &$ 8.31\substack{+0.01 \\ -0.01}$ &$ 8.31\substack{+0.02 \\ -0.02}$ &$ 8.23\substack{+0.02 \\ -0.02}$ \\
 9.0 & $ 7.95\substack{+0.00 \\ -0.00}$ &$ 8.10\substack{+0.00 \\ -0.00}$ &$ 8.27\substack{+0.01 \\ -0.01}$ &$ 8.14\substack{+0.02 \\ -0.02}$ &$ 8.18\substack{+0.01 \\ -0.01}$ &$ 8.19\substack{+0.02 \\ -0.02}$ &$ 8.09\substack{+0.02 \\ -0.02}$ \\
10.3 & $ 7.82\substack{+0.00 \\ -0.00}$ &$ 7.99\substack{+0.00 \\ -0.00}$ &$ 8.16\substack{+0.01 \\ -0.01}$ &$ 8.03\substack{+0.02 \\ -0.01}$ &$ 8.06\substack{+0.01 \\ -0.01}$ &$ 8.09\substack{+0.02 \\ -0.02}$ &$ 7.97\substack{+0.02 \\ -0.02}$ \\
11.7 & $ 7.70\substack{+0.00 \\ -0.00}$ &$ 7.88\substack{+0.00 \\ -0.00}$ &$ 8.06\substack{+0.01 \\ -0.01}$ &$ 7.93\substack{+0.02 \\ -0.02}$ &$ 7.96\substack{+0.01 \\ -0.01}$ &$ 7.99\substack{+0.02 \\ -0.02}$ &$ 7.85\substack{+0.02 \\ -0.02}$ \\
13.0 & $ 7.60\substack{+0.00 \\ -0.00}$ &$ 7.78\substack{+0.00 \\ -0.00}$ &$ 7.98\substack{+0.01 \\ -0.01}$ &$ 7.85\substack{+0.02 \\ -0.02}$ &$ 7.87\substack{+0.01 \\ -0.01}$ &$ 7.90\substack{+0.02 \\ -0.02}$ &$ 7.75\substack{+0.02 \\ -0.02}$ \\
14.5 & $ 7.50\substack{+0.00 \\ -0.00}$ &$ 7.69\substack{+0.00 \\ -0.00}$ &$ 7.90\substack{+0.01 \\ -0.01}$ &$ 7.76\substack{+0.02 \\ -0.02}$ &$ 7.78\substack{+0.01 \\ -0.01}$ &$ 7.82\substack{+0.02 \\ -0.02}$ &$ 7.65\substack{+0.02 \\ -0.02}$ \\
16.0 & $ 7.39\substack{+0.00 \\ -0.00}$ &$ 7.60\substack{+0.00 \\ -0.00}$ &$ 7.82\substack{+0.01 \\ -0.01}$ &$ 7.68\substack{+0.02 \\ -0.02}$ &$ 7.69\substack{+0.01 \\ -0.01}$ &$ 7.74\substack{+0.02 \\ -0.03}$ &$ 7.56\substack{+0.02 \\ -0.03}$ \\
17.3 & $ 7.31\substack{+0.00 \\ -0.00}$ &$ 7.52\substack{+0.00 \\ -0.00}$ &$ 7.76\substack{+0.01 \\ -0.01}$ &$ 7.61\substack{+0.02 \\ -0.02}$ &$ 7.62\substack{+0.01 \\ -0.01}$ &$ 7.67\substack{+0.03 \\ -0.03}$ &$ 7.48\substack{+0.03 \\ -0.03}$ \\
18.7 & $ 7.23\substack{+0.00 \\ -0.00}$ &$ 7.45\substack{+0.00 \\ -0.00}$ &$ 7.69\substack{+0.01 \\ -0.01}$ &$ 7.55\substack{+0.02 \\ -0.02}$ &$ 7.55\substack{+0.01 \\ -0.01}$ &$ 7.61\substack{+0.03 \\ -0.03}$ &$ 7.40\substack{+0.03 \\ -0.03}$ \\
22.6 & $ 7.02\substack{+0.00 \\ -0.00}$ &$ 7.27\substack{+0.00 \\ -0.00}$ &$ 7.54\substack{+0.01 \\ -0.01}$ &$ 7.38\substack{+0.02 \\ -0.02}$ &$ 7.37\substack{+0.01 \\ -0.01}$ &$ 7.45\substack{+0.03 \\ -0.03}$ &$ 7.21\substack{+0.03 \\ -0.03}$ \\
26.1 & $ 6.86\substack{+0.00 \\ -0.00}$ &$ 7.12\substack{+0.00 \\ -0.00}$ &$ 7.41\substack{+0.01 \\ -0.01}$ &$ 7.25\substack{+0.02 \\ -0.02}$ &$ 7.24\substack{+0.01 \\ -0.01}$ &$ 7.32\substack{+0.03 \\ -0.03}$ &$ 7.05\substack{+0.03 \\ -0.03}$ \\
30.0 & $ 6.70\substack{+0.00 \\ -0.00}$ &$ 6.98\substack{+0.00 \\ -0.00}$ &$ 7.29\substack{+0.01 \\ -0.01}$ &$ 7.13\substack{+0.03 \\ -0.02}$ &$ 7.10\substack{+0.01 \\ -0.01}$ &$ 7.20\substack{+0.03 \\ -0.04}$ &$ 6.90\substack{+0.03 \\ -0.04}$ \\
33.7 & $ 6.55\substack{+0.00 \\ -0.00}$ &$ 6.85\substack{+0.01 \\ -0.01}$ &$ 7.18\substack{+0.01 \\ -0.01}$ &$ 7.01\substack{+0.03 \\ -0.03}$ &$ 6.98\substack{+0.01 \\ -0.01}$ &$ 7.09\substack{+0.03 \\ -0.03}$ &$ 6.76\substack{+0.03 \\ -0.03}$ \\
37.8 & $ 6.41\substack{+0.00 \\ -0.00}$ &$ 6.72\substack{+0.01 \\ -0.01}$ &$ 7.07\substack{+0.01 \\ -0.01}$ &$ 6.90\substack{+0.03 \\ -0.03}$ &$ 6.85\substack{+0.01 \\ -0.01}$ &$ 6.98\substack{+0.04 \\ -0.04}$ &$ 6.63\substack{+0.04 \\ -0.04}$ \\
41.6 & $ 6.29\substack{+0.01 \\ -0.01}$ &$ 6.61\substack{+0.01 \\ -0.01}$ &$ 6.98\substack{+0.01 \\ -0.01}$ &$ 6.81\substack{+0.03 \\ -0.03}$ &$ 6.75\substack{+0.01 \\ -0.01}$ &$ 6.89\substack{+0.04 \\ -0.04}$ &$ 6.51\substack{+0.04 \\ -0.04}$ \\
45.7 & $ 6.17\substack{+0.01 \\ -0.01}$ &$ 6.50\substack{+0.01 \\ -0.01}$ &$ 6.88\substack{+0.01 \\ -0.01}$ &$ 6.71\substack{+0.03 \\ -0.03}$ &$ 6.64\substack{+0.01 \\ -0.01}$ &$ 6.79\substack{+0.04 \\ -0.04}$ &$ 6.39\substack{+0.04 \\ -0.04}$ \\
49.3 & $ 6.07\substack{+0.01 \\ -0.01}$ &$ 6.41\substack{+0.01 \\ -0.01}$ &$ 6.80\substack{+0.01 \\ -0.02}$ &$ 6.62\substack{+0.03 \\ -0.03}$ &$ 6.56\substack{+0.01 \\ -0.01}$ &$ 6.70\substack{+0.04 \\ -0.04}$ &$ 6.30\substack{+0.04 \\ -0.04}$ \\
53.1 & $ 5.98\substack{+0.01 \\ -0.01}$ &$ 6.33\substack{+0.01 \\ -0.01}$ &$ 6.71\substack{+0.02 \\ -0.02}$ &$ 6.55\substack{+0.03 \\ -0.03}$ &$ 6.46\substack{+0.01 \\ -0.01}$ &$ 6.64\substack{+0.04 \\ -0.04}$ &$ 6.21\substack{+0.04 \\ -0.04}$ \\
57.2 & $ 5.88\substack{+0.01 \\ -0.01}$ &$ 6.24\substack{+0.01 \\ -0.01}$ &$ 6.63\substack{+0.02 \\ -0.02}$ &$ 6.47\substack{+0.04 \\ -0.04}$ &$ 6.37\substack{+0.01 \\ -0.01}$ &$ 6.56\substack{+0.04 \\ -0.04}$ &$ 6.11\substack{+0.04 \\ -0.04}$ \\
61.5 & $ 5.79\substack{+0.01 \\ -0.01}$ &$ 6.15\substack{+0.01 \\ -0.01}$ &$ 6.55\substack{+0.02 \\ -0.02}$ &$ 6.39\substack{+0.04 \\ -0.04}$ &$ 6.29\substack{+0.01 \\ -0.01}$ &$ 6.49\substack{+0.04 \\ -0.04}$ &$ 6.03\substack{+0.04 \\ -0.04}$ \\
66.0 & $ 5.70\substack{+0.01 \\ -0.01}$ &$ 6.05\substack{+0.01 \\ -0.01}$ &$ 6.47\substack{+0.02 \\ -0.02}$ &$ 6.32\substack{+0.04 \\ -0.04}$ &$ 6.20\substack{+0.01 \\ -0.01}$ &$ 6.37\substack{+0.05 \\ -0.06}$ &$ 5.94\substack{+0.05 \\ -0.06}$ \\
69.8 & $ 5.64\substack{+0.01 \\ -0.01}$ &$ 5.98\substack{+0.01 \\ -0.01}$ &$ 6.40\substack{+0.02 \\ -0.02}$ &$ 6.25\substack{+0.04 \\ -0.04}$ &$ 6.12\substack{+0.02 \\ -0.01}$ &$ 6.35\substack{+0.04 \\ -0.05}$ &$ 5.87\substack{+0.04 \\ -0.05}$ \\
74.7 & $ 5.56\substack{+0.01 \\ -0.01}$ &$ 5.89\substack{+0.01 \\ -0.01}$ &$ 6.32\substack{+0.02 \\ -0.02}$ &$ 6.18\substack{+0.04 \\ -0.04}$ &$ 6.04\substack{+0.02 \\ -0.02}$ &$ 6.28\substack{+0.05 \\ -0.05}$ &$ 5.79\substack{+0.05 \\ -0.05}$ \\
79.9 & $ 5.49\substack{+0.01 \\ -0.01}$ &$ 5.81\substack{+0.01 \\ -0.01}$ &$ 6.24\substack{+0.02 \\ -0.02}$ &$ 6.12\substack{+0.04 \\ -0.04}$ &$ 5.96\substack{+0.02 \\ -0.02}$ &$ 6.20\substack{+0.05 \\ -0.06}$ &$ 5.72\substack{+0.05 \\ -0.06}$ \\
84.3 & $ 5.43\substack{+0.01 \\ -0.01}$ &$ 5.74\substack{+0.01 \\ -0.01}$ &$ 6.18\substack{+0.02 \\ -0.02}$ &$ 6.05\substack{+0.04 \\ -0.05}$ &$ 5.89\substack{+0.02 \\ -0.02}$ &$ 6.16\substack{+0.05 \\ -0.05}$ &$ 5.65\substack{+0.05 \\ -0.05}$ \\
88.8 & $ 5.38\substack{+0.01 \\ -0.01}$ &$ 5.67\substack{+0.01 \\ -0.01}$ &$ 6.11\substack{+0.02 \\ -0.02}$ &$ 5.99\substack{+0.05 \\ -0.06}$ &$ 5.81\substack{+0.02 \\ -0.02}$ &$ 6.08\substack{+0.05 \\ -0.06}$ &$ 5.58\substack{+0.05 \\ -0.06}$ \\
97.2 & $ 5.29\substack{+0.01 \\ -0.01}$ &$ 5.56\substack{+0.01 \\ -0.01}$ &$ 5.98\substack{+0.02 \\ -0.02}$ &$ 5.92\substack{+0.04 \\ -0.04}$ &$ 5.69\substack{+0.02 \\ -0.02}$ &$ 5.99\substack{+0.05 \\ -0.05}$ &$ 5.47\substack{+0.05 \\ -0.05}$ \\
103.6 & $ 5.21\substack{+0.01 \\ -0.01}$ &$ 5.49\substack{+0.01 \\ -0.01}$ &$ 5.89\substack{+0.03 \\ -0.03}$ &$ 5.84\substack{+0.05 \\ -0.05}$ &$ 5.62\substack{+0.02 \\ -0.02}$ &$ 5.94\substack{+0.05 \\ -0.05}$ &$ 5.39\substack{+0.05 \\ -0.05}$ \\
111.6 & $ 5.14\substack{+0.01 \\ -0.01}$ &$ 5.40\substack{+0.01 \\ -0.01}$ &$ 5.79\substack{+0.03 \\ -0.03}$ &$ 5.78\substack{+0.05 \\ -0.05}$ &$ 5.54\substack{+0.02 \\ -0.02}$ &$ 5.87\substack{+0.05 \\ -0.05}$ &$ 5.32\substack{+0.05 \\ -0.05}$ \\
117.2 & $ 5.10\substack{+0.01 \\ -0.01}$ &$ 5.36\substack{+0.01 \\ -0.01}$ &$ 5.72\substack{+0.03 \\ -0.03}$ &$ 5.72\substack{+0.05 \\ -0.05}$ &$ 5.47\substack{+0.02 \\ -0.02}$ &$ 5.82\substack{+0.05 \\ -0.05}$ &$ 5.29\substack{+0.05 \\ -0.05}$ \\
129.0 & $ 5.00\substack{+0.01 \\ -0.01}$ &$ 5.25\substack{+0.02 \\ -0.02}$ &$ 5.61\substack{+0.03 \\ -0.03}$ &$ 5.64\substack{+0.05 \\ -0.05}$ &$ 5.36\substack{+0.02 \\ -0.02}$ &$ 5.74\substack{+0.05 \\ -0.05}$ &$ 5.21\substack{+0.05 \\ -0.05}$ \\
141.7 & $ 4.89\substack{+0.02 \\ -0.02}$ &$ 5.13\substack{+0.02 \\ -0.02}$ &$ 5.49\substack{+0.03 \\ -0.03}$ &$ 5.58\substack{+0.05 \\ -0.05}$ &$ 5.23\substack{+0.03 \\ -0.03}$ &$ 5.66\substack{+0.05 \\ -0.05}$ &$ 5.09\substack{+0.05 \\ -0.05}$ \\
146.7 & $ 4.85\substack{+0.02 \\ -0.02}$ &$ 5.10\substack{+0.02 \\ -0.02}$ &$ 5.46\substack{+0.03 \\ -0.03}$ &$ 5.51\substack{+0.06 \\ -0.06}$ &$ 5.19\substack{+0.03 \\ -0.03}$ &$ 5.61\substack{+0.05 \\ -0.05}$ &$ 5.03\substack{+0.05 \\ -0.05}$ \\

\enddata
\tablecomments{
    Average \mden{} profiles of massive \rbcg{} and \nbcg{} galaxies in different
    samples:\\ 
    Col.~(1) Radius along the major axis in kpc.\\
    Col.~(2) Average \mden{} profile for galaxies with 
        $11.4 \leq$\logmtot$< 11.6$ in the combined samples of \rbcg{} and \nbcg{}
        galaxies. \\ 
    Col.~(3) Average \mden{} profile of combined samples in the mass bin of 
        $11.6 \leq$\logmtot$< 11.8$. \\ 
    Col.~(4) Average \mden{} profile of combined samples in the mass bin of 
        $11.8 \leq$\logmtot$< 12.0$. \\ 
    Col.~(5) and Col.~(6) are the average \mden{} profiles of \rbcg{} and \nbcg{} galaxies
        in the \mtot{}-matched samples within $11.6 \leq$\logmtot{}$< 11.9$. \\ 
    Col.~(7) and Col.~(8) are the average \mden{} profiles of \rbcg{} and \nbcg{} galaxies 
        in the \minn{}-matched samples within $11.2 \leq$\logmtot{}$< 11.6$. \\ 
    The upper and lower uncertainties of these average profiles vial bootstrap-resampling 
    method are also displayed.
}
\label{tab:prof}
\end{deluxetable}